\documentclass{article}
\pdfoutput=1

\usepackage[utf8]{inputenc}
\usepackage{lmodern}
\usepackage{hyperref}
\usepackage[super]{nth}
\makeatletter
\def\namedlabel#1#2{\begingroup
    #2%
    \def\@currentlabel{#2}%
    \phantomsection\label{#1}\endgroup
}
\makeatother

\usepackage{graphicx}       
\usepackage{subcaption}
\usepackage{tikz}
\usetikzlibrary{shapes.geometric, arrows, positioning, calc}

\usepackage[acronym,section,nonumberlist]{glossaries}
\makenoidxglossaries
\newacronym{pl}{PL}{Parameter Logistic}
\newacronym{3pl}{3PL}{Three-Parameter Logistic}
\newacronym{4pl}{4PL}{Four-Parameter Logistic}
\newacronym{ctt}{CTT}{Classical Test Theory}
\newacronym{dif}{DIF}{Differential Item Functioning}
\newacronym{em}{EM}{Expectation-Maximisation} 
\newacronym{glnm}{GLNMs}{Generalized Linear and Nonlinear Models}
\newacronym{irt}{IRT}{Item Response Theory}
\newacronym{ml}{ML}{Maximum Likelihood}
\newacronym{nls}{NLS}{Nonlinear Least Squares}
\newacronym{plf}{PLF}{Parametrized Link Function}
\newacronym{rss}{RSS}{Residual Sum of Squares}
\newacronym{rmse}{RMSE}{Root Mean Squared Error}

\usepackage{amsmath}
\usepackage{amssymb}
\usepackage{mathbbol}

\DeclareMathOperator{\E}{\textsf{E}}

\newcommand{\goto}{\rightarrow}

\newcommand{\bgamma}{\boldsymbol{\gamma}_i}
\newcommand{\bgammahat}{\hat{\boldsymbol{\gamma}}_i}
\newcommand{\bgammaX}{\boldsymbol{\gamma}_{iX}}
\newcommand{\T}[1]{#1^\top}
\usepackage{stackrel}
\newcommand{\gotop}{\stackrel[n \goto \infty]{P}{\longrightarrow}}
\newcommand{\gotod}{\stackrel[n \goto \infty]{\mathcal{D}}{\longrightarrow}}

\usepackage{tabularx}
\usepackage{booktabs}
\usepackage{adjustbox}
\usepackage{longtable}
\usepackage{siunitx}
\usepackage{multirow}
\usepackage{threeparttable}

\newcommand{\tablenotet}[1]{%
\hphantom{x}\\[-2.25ex]
{\raggedleft\small%
\textit{Note. }{#1}%
}}

\usepackage{array}
\newcommand{\PreserveBackslash}[1]{\let\temp=\\#1\let\\=\temp}
\newcolumntype{C}[1]{>{\PreserveBackslash\centering}p{#1}}
\newcolumntype{R}[1]{>{\PreserveBackslash\raggedleft}p{#1}}

\DeclareRobustCommand\Rcode{\bgroup\@codex}
\def\@codex#1{\texorpdfstring%
{{\normalfont\ttfamily #1}}%
{#1}\egroup}

\let\pkg=\strong
\usepackage{apacite}

\newcommand{\highlight}[1]{\textit{#1}}
\usepackage{comment}

\usepackage{appendix}

\usepackage{arxiv}

\title{New iterative algorithms for estimation of item functioning}

\date{}

\author{Adéla Hladká$^{1,2}$, Patrícia Martinková$^{1,3}$, Marek Brabec$^{1}$\\
\small $^{1}$ Institute of Computer Science of the Czech Academy of Sciences, Prague, Czech Republic\\
\small $^{2}$ Faculty of Mathematics and Physics, Charles University, Prague, Czech Republic\\
\small $^{3}$ Faculty of Education, Charles University, Prague, Czech Republic\\
}

\begin{document}

\maketitle


\begin{abstract}
This paper explores innovations to parameter estimation in generalized linear and nonlinear models, which may be used in item response modeling to account for guessing/pretending or slipping/dissimulation and for the effect of covariates. We introduce a new implementation of the EM algorithm and propose a new algorithm based on the parametrized link function. The two novel iterative algorithms are compared to existing methods in a simulation study. Additionally, the study examines software implementation, including the specification of initial values for numerical algorithms and asymptotic properties with an estimation of standard errors. Overall, the newly proposed algorithm based on the parametrized link function outperforms other procedures, especially for small sample sizes. Moreover, the newly implemented EM algorithm provides additional information regarding respondents' inclination to guess or pretend and slip or dissimulate when answering the item. The study also discusses applications of the methods in the context of the detection of differential item functioning and addresses the measurement error. Methods are offered in the \pkg{difNLR} package and in the interactive application of the \pkg{ShinyItemAnalysis} package; demonstration is provided using real data from psychological and educational assessments.  
\end{abstract}


\section{Introduction}

In fields such as education, psychology, and health, constructs are commonly measured through multi-item instruments, where understanding how each individual item functions is crucial. Analyzing item functioning not only aids in refining measurement instruments but also provides valuable insights into the behavior and characteristics of different respondent groups. The \gls{3pl} and \gls{4pl} models are flexible tools that allow capturing complex item response patterns and accommodating a more comprehensive range of item characteristics, including possible guessing and slipping rates in the context of educational measurement and pretending and dissimulation in the context of psychological and health measurement. However, estimation in these models, both in the \gls{irt} \cite{birnbaum1968statistical, barton1981upper} and non-\gls{irt} framework \cite{drabinova2017detection, hladka2020difnlr}, may become challenging due to several factors, including the complexity of these models caused by their non-linearity, high-dimensionality of the parameter space, and the nature of the data being analyzed. These models typically require a large sample size \cite{kim2013effect}, which can result in computationally demanding fitting. Therefore, efficient algorithms, advanced estimation techniques, and software implementation are crucial for the effectiveness and accessibility of these models' use in practice. 

Recent research renewed interest in the 3--\gls{4pl} \gls{irt} models since the availability of computing resources is on the rise. New approaches in estimation are being studied extensively \cite{battauz2020regularized, culpepper2016revisiting, loken2010estimation, meng2020marginalized, fu2021gibbs}, which help solve some of the computational issues. However, their focus is mainly limited to large-scale assessments, while estimating item parameters with moderate sample sizes is still unreachable. To address the computational issues more effectively and accurately recover item characteristics such as guessing and slipping, the \gls{irt} models may benefit from traditional item analysis and \gls{glnm}, their simpler score-based counterparts \cite{martinkova2023computational}. The \gls{glnm} can offer improved and more precise starting values for related \gls{irt} models and allow for statistical inference regarding item parameters while still being less computationally complex compared to \gls{irt} models. 

\gls{glnm} incorporates a class of generalized logistic regression models that are natural extensions of the logistic regression model to describe item functioning. Analogously to 3--\gls{4pl} \gls{irt} models, generalized logistic regression may account for the possibility that an item can be correctly answered or endorsed without the necessary knowledge or trait, e.g., due to guessing or pretending. In this case, the logistic regression model is extended by including a parameter defining a lower asymptote of the probability curve, which may be larger than zero. Similarly, the model can consider the possibility that an item is incorrectly answered or opposed by a respondent with a high level of a particular trait due to issues such as inattention, lack of time, or dissimulation; this model includes an upper asymptote of the probability curve, which may be lower than one. These models can be seen as score-based counterparts to 3--\gls{4pl} \gls{irt} models since they assume the same shape of the item response curve; however, in contrast to the class of latent variable models, this approach uses an observed estimate of the underlying latent trait. 

Furthermore, logistic regression, its extensions, and their latent variable counterparts have become widely used for identifying between-group differences on item level when responding to multi-item measurements \cite{swaminathan1990detecting}. The phenomenon, known as \gls{dif}, indicates whether responses to an item vary for respondents with the same level of an underlying latent trait but from different groups (e.g., defined by gender, age, or socio-economic status). In this vein, \gls{dif} detection is essential for a deeper understanding of group differences, assessing the effectiveness of various treatments, or uncovering potential unfairness in educational tests. It is identified as one of the crucial topics in measurement (AERA, APA, \& NCME, \citeyearNP{aera2014standards}). 

The estimation in the logistic regression model is a straightforward procedure, but extending the parametric space by including additional parameters in this model makes it more statistically and computationally challenging and demanding and may result in convergence issues. This is even more present in \gls{irt} modeling, where latent ability is estimated together with item parameters. In this vein, \gls{glnm} can be seen as a helpful alternative in describing item functioning and identifying \gls{dif}, accounting for possible guessing or inattention while also being accessible in practice.

Therefore, this article examines innovations in the item parameter estimation for the \gls{glnm} in the context of \gls{dif} detection. As the main contribution, the work proposes novel iterative algorithms, examines their theoretical properties, and compares the newly proposed methods to existing ones in a simulation study. The use of estimation procedures is then exemplified on real data examples with an application to \gls{dif} detection, with the secondary goal of providing possibilities for more accurate \gls{dif} detection. Given that GLNMs treat ability as observed, we also address potential biases this approach may bring due to measurement error.

The rest of the manuscript is organized as follows: To begin, Section \ref{sec:methodology} introduces the \gls{glnm} and its relationship to \gls{irt} framework, examining the estimation techniques. This section provides a detailed description of two existing methods for parameter estimation, the \gls{nls} and the \gls{ml} method, and their application to fitting \gls{glnm}. Furthermore, as an alternative to the direct implementation of the \gls{ml} method, this study proposes a novel implementation of the \gls{em} algorithm and a new approach based on a \gls{plf}. Additionally, this section provides asymptotic properties of the estimates, an estimation of standard errors, and a software implementation, including a specification of starting values in iterative algorithms. Subsequently, Section~\ref{sec:simulation} describes the design and results of the simulation study. To illustrate differences and challenges between the existing and newly proposed methods in practice and the context of \gls{dif} detection, this work provides two real data analyses in Section \ref{sec:real-data}. Section \ref{sec:discussion} contains the discussion and concluding remarks. Finally, Appendix \ref{app:sec:asymptotics} provides asymptotic properties of the discussed estimation approaches, Appendix \ref{app:sec:tables} lists item parameter estimates of real data examples, and Appendix \ref{app:sec:error} presents an additional study on measurement error.


\section{Methodology}\label{sec:methodology}


\subsection{Generalized linear and nonlinear models for item functioning}

\gls{glnm} extend the logistic regression model by accounting for the possibility of guessing or inattention when answering an item. The \highlight{simple \gls{4pl} model} describes functioning of the item $i$, meaning the probability of endorsing item $i$ by respondent $p$, by introducing four parameters: 
\begin{align}\label{eq:4PL:simple}
    \pi_{pi} = \mathrm{P}(Y_{pi} = 1|\theta_p) = c_i + (d_i - c_i)\ \frac{\exp(b_{i0} + b_{i1} \theta_p)}{1 + \exp(b_{i0} + b_{i1} \theta_p)},
\end{align}
with $\theta_p$ being an observed trait of respondent $p$. 

\paragraph{Parameter interpretation. }
All four parameters have an intuitive interpretation: The parameters $c_i$ and $d_i$ are the lower and upper asymptotes of the probability sigmoid function $\pi_{pi}(x)$ since
\begin{align*}
    \lim_{x \to -\infty}\pi_{pi}(x) = c_i \quad \text{and} \quad
    \lim_{x \to \infty}\pi_{pi}(x) = d_i,
\end{align*}
where $c_i \in [0, 1], d_i \in [0, 1]$ and $c_i < d_i$ if $b_{i1} > 0$ and $c_i > d_i$ otherwise. Evidently, with $c_i = 0$ and $d_i = 1$, this model recovers a standard logistic regression for item $i$.

In psychological and health-related assessments, the asymptotes $c_i$ may represent pretending or simulation, and $1 - d_i$ may represent the probability of reluctance to admit difficulties due to social norms or dissimulation. In educational testing, parameter $c_i$ can be interpreted as the probability that the respondents guessed the correct answer without possessing the necessary knowledge $\theta_p$, also known as a pseudo-guessing parameter. On the other hand, $1 - d_i$ can be viewed as the probability that respondents were inattentive while their knowledge $\theta_p$ was sufficient \cite{hladka2020difnlr}, or a lapse-rate \cite{kingdom2016psychophysics}. Next, the parameter $b_{i0}$ is an intercept parameter related to the difficulty of item $i$ (or item popularity in psychological and health-related assessments), and parameter $b_{i1}$ is linked to a slope of the sigmoid curve $\pi_{pi}(x)$, which is also called discrimination of the respective item.

\paragraph{Adding covariates, group-specific \gls{4pl} model. }
The simple model \eqref{eq:4PL:simple} can be further extended by incorporating additional respondents' characteristics. As a typical example, a binary grouping variable $G_p$ might be considered. This variable describes a respondent's membership to a social group ($G_p = 0$ for the reference group and $G_p = 1$ for the focal group), which extends the simple \gls{4pl} model~\eqref{eq:4PL:simple} to a group-specific form:
\begin{align}\label{eq:4PL:dif}
    \begin{split}
        \pi_{pi} =\ &\mathrm{P}(Y_{pi} =\ 1|\theta_p, G_p) = c_i + c_{i\text{DIF}}G_p\\
        & + (d_i - d_{i\text{DIF}}G_p - c_i - c_{i\text{DIF}}G_p)\ \frac{\exp(b_{i0} + b_{i1} \theta_p + b_{i2}G_p + b_{i3}\theta_p\cdot G_p)}{1 + \exp(b_{i0} + b_{i1} \theta_p + b_{i2}G_p + b_{i3}\theta_p\cdot G_p)},
    \end{split}
\end{align}
which is suitable for testing \gls{dif}, see also \citeA{hladka2020difnlr} and \citeA[Chapter 9]{martinkova2023computational}. 

However, the models \eqref{eq:4PL:simple} and \eqref{eq:4PL:dif} can be further generalized. Instead of using a single variable $\theta_p$ to describe item functioning or added grouping variable $G_p$ to test for \gls{dif}, we introduce in this paper a vector of covariates $\boldsymbol{X}_p = (1,  X_{p1}, \dots, X_{pk})^\intercal$, $p = 1, \dots, n$, which includes the original observed trait and an intercept term. This process produces extra parameters $\boldsymbol{b}_i = (b_{i0}, \dots, b_{ik})^\intercal$. Beyond this, even asymptotes may depend on respondents' characteristics $\boldsymbol{Z}_p = (1, Z_{p1}, \dots, Z_{pj})^\intercal$, $p = 1, \dots, n$, which are not necessarily the same as $\boldsymbol{X}_p$. This general \highlight{covariate-specific \gls{4pl} model} is of form 
\begin{align}\label{eq:4PL:gen}
    \pi_{pi} = \mathrm{P}(Y_{pi} = 1|\boldsymbol{X}_p, \boldsymbol{Z}_p) = \boldsymbol{Z}_p^\intercal\boldsymbol{c}_i + (\boldsymbol{Z}_p^\intercal\boldsymbol{d}_i - \boldsymbol{Z}_p^\intercal\boldsymbol{c}_i)\ \frac{\exp(\boldsymbol{X}_p^\intercal\boldsymbol{b}_i)}{1 + \exp(\boldsymbol{X}_p^\intercal\boldsymbol{b}_i)},
\end{align}
where $\boldsymbol{c}_i = (c_{i0}, \dots, c_{ij})^\intercal$ and $\boldsymbol{d}_i = (d_{i0}, \dots, d_{ij})^\intercal$ are asymptote parameters for item $i$. Note that we typically assume $\boldsymbol{Z}_p$ being categorical rather than continuous variables describing respondents' characteristics to keep meaningful interpretation while requiring $0 \leq \boldsymbol{Z}_p^\intercal\boldsymbol{c}_i < \boldsymbol{Z}_p^\intercal\boldsymbol{d}_i \leq 1$. With $\boldsymbol{X}_p = (1, \theta_{p})$ and $\boldsymbol{Z}_p = 1$, we get the simple \gls{4pl} model \eqref{eq:4PL:simple}, while  the choice of $\boldsymbol{X}_p = (1, \theta_{p}, G_p)$ and $\boldsymbol{Z}_p = (1, G_p)$ yields the group-specific \gls{4pl} model \eqref{eq:4PL:dif}.

\paragraph{Testing for \gls{dif}. }
The \highlight{group-specific \gls{4pl} model}~\eqref{eq:4PL:dif} can be then used for testing between-group differences on the item level with a \gls{dif} analysis \cite{hladka2020difnlr} with, for example, the likelihood-ratio test. The likelihood-ratio test measures the difference between the log-likelihood $l_{i1}$ of the larger model (e.g., the group-specific \gls{4pl} model~\eqref{eq:4PL:dif}) and the log-likelihood $l_{i0}$ of its submodel (e.g., the simple \gls{4pl} model~\eqref{eq:4PL:simple}) for the given item $i$. The resulting $LR_i$ statistic has an asymptotic $\chi^2$-distribution under the  smaller model with degrees of freedom equal to a difference in the number of parameters in the two models:
\begin{align}\label{eq:lrt}
    LR_i = -2 \left(l_{i0} - l_{i1}\right) \gotod \chi^2(\text{df}_{i1} - \text{df}_{i0}).
\end{align}
Similarly, any two nested submodels of the group-specific \gls{4pl} model~\eqref{eq:4PL:dif} can be compared to test for the significance of group-related item parameters. Note that in \gls{glnm}, \gls{dif} detection is usually performed item-by-item.

\paragraph{Matching criterion. }
In these models, $\theta_p$ is an observed variable describing the measured trait of the respondent, such as anxiety, fatigue, quality of life, or math ability, here called the \highlight{matching criterion}. In the context of the logistic regression method for \gls{dif} detection, the total test score (or its standardized version) is typically used as the matching criterion \cite{swaminathan1990detecting}. Other options for the matching criterion include a pre-test score (to identify differential item functioning in change; see \citeauthor{martinkova2020academic}, 
\citeyearNP{martinkova2020academic}), a score on another test measuring the same construct, or an estimate of the latent trait provided by an \gls{irt} model. 

\paragraph{IRT framework. }
Within the \gls{irt} framework, the matching criterion $\theta_p$ in the models \eqref{eq:4PL:simple}--\eqref{eq:4PL:dif} is replaced by a latent ability $\theta$, necessitating joint estimation with item parameters. Both frameworks share the same shape of item characteristic curves with a comparable interpretation. A notable distinction lies in the estimation process: \gls{irt} models simultaneously estimate parameters for all items, which is typically not the case for models \eqref{eq:4PL:simple}--\eqref{eq:4PL:dif} as described below. However, \gls{glnm} entail lower computational demands, as they require smaller sample sizes to yield precise estimates. The estimation algorithms for \gls{glnm}, which are the focus of this paper, may thus be further incorporated into the \gls{irt} framework as is further described in the Discussion. 


\subsection{Estimation of item parameters}

Numerous algorithms are available to estimate item parameters in the covariate-specific \gls{4pl} model~\eqref{eq:4PL:gen}. First, this section describes two methods that may be directly implemented in the existing software: The \gls{nls} method and the \gls{ml} method. Next, the study introduces two newly proposed iterative algorithms, which might improve the implementation of the computationally demanding \gls{ml} method: The \gls{em} algorithm inspired by the work of \citeA{dinse2011algorithm} and an iterative algorithm based on \gls{plf}. The described and proposed algorithms treat ability as known or estimated apriori and estimate parameters for each item separately, which is suitable in the GLNM framework. We provide a discussion on incorporating joint ability and item parameter estimation in the Discussion section. 


\subsubsection{Nonlinear least squares}

The parameter estimates of the covariate-specific \gls{4pl} model~\eqref{eq:4PL:gen} can be determined using the \gls{nls} method \cite{dennis1981adaptive, drabinova2017detection, hladka2020difnlr}, which is based on minimisation of the \gls{rss} of item $i$ with respect to item parameters $(\boldsymbol{b}_i, \boldsymbol{c}_i, \boldsymbol{d}_i)$:
\begin{align}\label{eq:nls:rss}
	\text{RSS}_i(\boldsymbol{b}_i, \boldsymbol{c}_i, \boldsymbol{d}_i) = \sum_{p = 1}^n \left[Y_{pi} - \pi_{pi}\right]^2  = \sum_{p = 1}^n \left[Y_{pi} - \boldsymbol{Z}_p^\intercal\boldsymbol{c}_i - (\boldsymbol{Z}_p^\intercal\boldsymbol{d}_i - \boldsymbol{Z}_p^\intercal\boldsymbol{c}_i)\frac{\exp(\boldsymbol{X}_p^\intercal\boldsymbol{b}_i)}{1 + \exp(\boldsymbol{X}_p^\intercal\boldsymbol{b}_i)}\right]^2,
\end{align}
where $n$ is the number of respondents. Since the criterion function $\text{RSS}_i(\boldsymbol{b}_i, \boldsymbol{c}_i, \boldsymbol{d}_i)$ is continuously differentiable with respect to item parameters $(\boldsymbol{b}_i, \boldsymbol{c}_i, \boldsymbol{d}_i)$, the minimiser can be obtained when the gradient is zero. Thus, the minimization process involves a calculation of the first partial derivatives with respect to item parameters $(\boldsymbol{b}_i, \boldsymbol{c}_i, \boldsymbol{d}_i)$ and finding a solution of relevant nonlinear estimating equations \cite<e.g., >[Chapter 5]{van2000asymptotic}. Since $\boldsymbol{Z}_p^\intercal\boldsymbol{c}_i$ and $\boldsymbol{Z}_p^\intercal\boldsymbol{d}_i$ asymptotes represent probabilities, it is necessary to ensure that these expressions are kept in the interval of $[0, 1]$ which is accomplished using numerical approaches. 

The asymptotic properties of the \gls{nls} estimator, such as consistency and asymptotic distribution, can be derived under the classical set of regularity conditions \cite<e.g.,>[Theorems 5.41 and 5.42; see also Appendix \ref{app:asymptotics:nls} for more details]{van2000asymptotic}. This study proposes a sandwich estimator \eqref{app:eq:nls:sandwich}, which can be used as a natural estimate of the asymptotic variance of the \gls{nls} estimate. 


\subsubsection{Maximum likelihood}

The second option for estimating item parameters in the covariate-specific \gls{4pl} model \eqref{eq:4PL:gen} is the \gls{ml} method \cite{hladka2020difnlr}. Using a notation $\phi_{pi} = \frac{\exp(\boldsymbol{X}_p^\intercal\boldsymbol{b}_i)}{1 + \exp(\boldsymbol{X}_p^\intercal\boldsymbol{b}_i)}$, the corresponding likelihood function for item $i$ has the following form:
\begin{align*}
    L_i(\boldsymbol{b}_i, \boldsymbol{c}_i, \boldsymbol{d}_i) = \prod_{p = 1}^n & \left[\boldsymbol{Z}_p^\intercal\boldsymbol{c}_i + (\boldsymbol{Z}_p^\intercal\boldsymbol{d}_i - \boldsymbol{Z}_p^\intercal\boldsymbol{c}_i)\phi_{pi}\right]^{Y_{pi}} \left[1 - \boldsymbol{Z}_p^\intercal\boldsymbol{c}_i - (\boldsymbol{Z}_p^\intercal\boldsymbol{d}_i - \boldsymbol{Z}_p^\intercal\boldsymbol{c}_i)\phi_{pi}\right]^{1 - Y_{pi}},
\end{align*}
and the log-likelihood function is then given by 
\begin{align*}
    l_{i}(\boldsymbol{b}_i, \boldsymbol{c}_i, \boldsymbol{d}_i) = \sum_{p = 1}^n &\left\{Y_{pi} \log(\boldsymbol{Z}_p^\intercal\boldsymbol{c}_i + (\boldsymbol{Z}_p^\intercal\boldsymbol{d}_i - \boldsymbol{Z}_p^\intercal\boldsymbol{c}_i)\phi_{pi})\right. \\
    &\ \left. +\ (1 - Y_{pi})\log(1 - \boldsymbol{Z}_p^\intercal\boldsymbol{c}_i - (\boldsymbol{Z}_p^\intercal\boldsymbol{d}_i - \boldsymbol{Z}_p^\intercal\boldsymbol{c}_i)\phi_{pi})\right\}.
\end{align*}

The parameter estimates are obtained by maximization of the log-likelihood function. Thus, this approach proceeds similarly to the logistic regression model, except for a larger dimension of the parametric space. To find the maximizer of the log-likelihood function $l_{i}(\boldsymbol{b}_i, \boldsymbol{c}_i, \boldsymbol{d}_i)$, the first partial derivatives are set to zero and these so-called likelihood equations must be solved. However, the solution of a system of nonlinear equations cannot be derived algebraically and needs to be numerically estimated using a suitable iterative process. 

Using \citeauthor{van2000asymptotic}'s \citeyear{van2000asymptotic} Theorems 5.41 and 5.42, consistency and asymptotic normality can be shown for the \gls{ml} estimator, see Appendix \ref{app:asymptotics:ml} for more details. Additionally, the estimate of the asymptotic variance of the item parameters is an inverse of the observed information matrix \eqref{app:eq:mle:variance}. 


\subsubsection{EM algorithm}

The \gls{ml} method may be computationally demanding, and iterative algorithms might help in those situations. Inspired by the work of \citeA{dinse2011algorithm}, this study adopts a version of the \gls{em} algorithm \cite{dempster1977maximum} for parameter estimation in the covariate-specific \gls{4pl} model \eqref{eq:4PL:gen}.

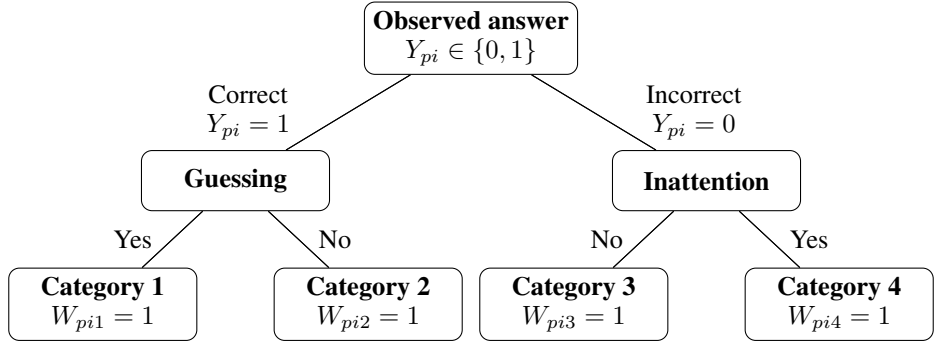
\begin{figure}[ht]
    \centering
    \begin{tikzpicture}
      \node [shape=rectangle, rounded corners, draw, align=center, minimum width=25mm, minimum height=8mm] (answer) {\textbf{Observed answer} \\ $Y_{pi} \in \left\{0, 1\right\}$}
        child { node [shape=rectangle, rounded corners, draw, align=center, minimum width=25mm, minimum height=8mm, below left=1cm and 0.45cm of answer](guessed) {\textbf{Guessing}} 
          child { node [shape=rectangle, rounded corners, draw, align=center, minimum width=25mm, minimum height=8mm, below left=0.75cm and -0.75cm of guessed](cat1) {\textbf{Category 1} \\ $W_{pi1} = 1$} }
          child { node [shape=rectangle, rounded corners, draw, align=center, minimum width=25mm, minimum height=8mm, below right=0.75cm and -0.75cm of guessed](cat2) {\textbf{Category 2} \\ $W_{pi2} = 1$} } }
        child { node [shape=rectangle, rounded corners, draw, align=center, minimum width=25mm, minimum height=8mm, below right=1cm and 0.45cm of answer](inattentive) {\textbf{Inattention}} 
          child { node [shape=rectangle, rounded corners, draw, align=center, minimum width=25mm, minimum height=8mm, below left=0.75cm and -0.75cm of inattentive](cat3) {\textbf{Category 3} \\ $W_{pi3} = 1$} }
          child { node [shape=rectangle, rounded corners, draw, align=center, minimum width=25mm, minimum height=8mm, below right=0.75cm and -0.75cm of inattentive](cat4) {\textbf{Category 4} \\ $W_{pi4} = 1$} } };
    \draw (answer) -- (guessed) node [shape=rectangle, rounded corners, align=center, minimum width=25mm, minimum height=8mm, midway, left, draw=none] {Correct \, \\ $Y_{pi} = 1$ \,};
    \draw (answer) -- (inattentive) node [shape=rectangle, rounded corners, align=center, minimum width=25mm, minimum height=8mm, midway, right, draw=none] {\, Incorrect \\ \, $Y_{pi} = 0$};
    \draw (guessed) -- (cat1) node [midway, left, draw=none] {Yes \,};
    \draw (guessed) -- (cat2) node [midway, right, draw=none] {\, No};
    \draw (inattentive) -- (cat3) node [midway, left, draw=none] {No \,};
    \draw (inattentive) -- (cat4) node [midway, right, draw=none] {\, Yes};
    \end{tikzpicture}
    \caption{Graphical representation of the relationships among latent variables for the \gls{em} algorithm}
    \label{fig:EM_latent_variables}
\end{figure}

To make use of the EM algorithm, the original 4PL model can be reformulated as a mixture model employing latent classes indicating different types of respondents. 
In our setting, we consider four mutually exclusive latent variables ($W_{pi1}$, $W_{pi2}$, $W_{pi3}$, $W_{pi4}$), where variable $W_{pij} = 1$ indicates that respondent $p$ belongs in the category $j = 1, \dots, 4$ for an item $i$, whereas $W_{pij} = 0$ indicates that respondent does not belong in this category. 
In the context of educational, psychological, health-related, or other types of multi-item measurement, the four categories can be interpreted as follows: Categories 1 and 2 indicate whether a respondent who responded correctly to item $i$ or endorsed it (i.e., $Y_{pi} = 1$) was determined to do so ($W_{pi1} = 1$, e.g., the respondent guessed correct answer while their knowledge or ability was insufficient, or the respondent simulated described situation) or not ($W_{pi2} = 1$, e.g., had sufficient knowledge or ability to answer correctly and did not guess, or endorsed while experiencing described situation). On the other hand, Categories 3 and 4 indicate whether the respondent who did not respond correctly or did not endorse the item (i.e., $Y_{pi} = 0$) was prone to do so ($W_{pi3} = 1$, e.g., did not have sufficient knowledge, ability, or trait) or not ($W_{pi4} = 1$, e.g., incorrectly answered, or did not endorse, due to another reason such as inattention, lack of time, or dissimulation). Thus, the observed indicator $Y_{pi}$ and its complement $1 - Y_{pi}$ could be rewritten as $Y_{pi} = W_{pi1} + W_{pi2}$ and $1 - Y_{pi} = W_{pi3} + W_{pi4}$ (Figure \ref{fig:EM_latent_variables}). While the latent variables $W_{pi2}$ and $W_{pi3}$ represent desired response styles, $W_{pi1}$ and $W_{pi4}$ represent response styles which are undesired. 

Let $\boldsymbol{Z}_p^\intercal\boldsymbol{c}_i$ be the regressor-based probability that the respondent was determined to respond to item $i$ correctly or endorse it (Category 1), and let $\boldsymbol{Z}_p^\intercal\boldsymbol{d}_i$ be the regressor-based probability of the respondent not prone to respond correctly or endorse item $i$ (Categories 1--3). Then $\boldsymbol{Z}_p^\intercal\boldsymbol{d}_i - \boldsymbol{Z}_p^\intercal\boldsymbol{c}_i$ gives the regressor-based probability that the respondent was not determined but prone to (Categories 2 and 3). Further, we denote $\phi_{pi}$ and $1 - \phi_{pi}$ -- the probabilities to answer a given item correctly (Category 2) and incorrectly (Category 3), respectively, depending on the regressors $\boldsymbol{X}_p$. Finally, the probability that the respondent did not respond correctly and was not prone to do so is given by $1 - (\boldsymbol{Z}_p^\intercal\boldsymbol{d}_i - \boldsymbol{Z}_p^\intercal\boldsymbol{c}_i) - \boldsymbol{Z}_p^\intercal\boldsymbol{c}_i = 1 - \boldsymbol{Z}_p^\intercal\boldsymbol{d}_i$ (Category 4). In summary, the expected values of the latent variables are then given by the following terms  
\begin{align*}
    \boldsymbol{Z}_p^\intercal\boldsymbol{c}_i, \ \ (\boldsymbol{Z}_p^\intercal\boldsymbol{d}_i - \boldsymbol{Z}_p^\intercal\boldsymbol{c}_i) \phi_{pi}, \ \ (\boldsymbol{Z}_p^\intercal\boldsymbol{d}_i - \boldsymbol{Z}_p^\intercal\boldsymbol{c}_i)(1 - \phi_{pi}), \ \ 1 - \boldsymbol{Z}_p^\intercal\boldsymbol{d}_i,
\end{align*}
and the probability of a correct response or endorsement is given by
\begin{align*}
    \mathrm{P}(Y_{pi} = 1|\boldsymbol{X}_p) =& \ \mathrm{P}(W_{pi1} + W_{pi2} =  1|\boldsymbol{X}_p)
    = \mathrm{P}(W_{pi1} = 1|\boldsymbol{X}_p) + \mathrm{P}(W_{pi2} = 1|\boldsymbol{X}_p) \\
    =&\ \boldsymbol{Z}_p^\intercal\boldsymbol{c}_i + (\boldsymbol{Z}_p^\intercal\boldsymbol{d}_i - \boldsymbol{Z}_p^\intercal\boldsymbol{c}_i)\phi_{pi},
\end{align*}
which under the logistic model $\phi_{pi} = \frac{\exp(\boldsymbol{X}_p^\intercal\boldsymbol{b}_i)}{1 + \exp(\boldsymbol{X}_p^\intercal\boldsymbol{b}_i)}$ produces the covariate-specific \gls{4pl} model \eqref{eq:4PL:gen}.

Using the setting of the latent variables, the corresponding log-likelihood function for~item $i$ takes the following form:
\begin{align*}
    l_i^{\text{EM}}
       =& \sum_{p = 1}^n \left[W_{pi2}\log\left(\phi_{pi}\right) + W_{pi3}\log\left(1 - \phi_{pi}\right)\right] \\
       &  + \sum_{p = 1}^n \left[W_{pi1}  \log\left(\boldsymbol{Z}_p^\intercal\boldsymbol{c}_i\right) + W_{pi4}\log\left(1 - \boldsymbol{Z}_p^\intercal\boldsymbol{d}_i\right) + \ \left(W_{pi2} + W_{pi3}\right) \log\left(\boldsymbol{Z}_p^\intercal\boldsymbol{d}_i - \boldsymbol{Z}_p^\intercal\boldsymbol{c}_i\right)\right] \\
    =& \ l_{i1}^{\text{EM}} + l_{i2}^{\text{EM}}.
\end{align*}
The log-likelihood function $l_{i1}^{\text{EM}}$ includes only parameters $\boldsymbol{b}_i$ and regressors $\boldsymbol{X}_p$, whereas the log-likelihood function $l_{i2}^{\text{EM}}$ incorporates only parameters related to the asymptotes of the sigmoid function and includes only regressors $\boldsymbol{Z}_p$. Notably, the log-likelihood function $l_{i1}^{\text{EM}}$ has a form of the log-likelihood function for the logistic regression. However, in contrast to the logistic regression model, in this setting, it does not necessarily hold that $W_{pi2} + W_{pi3} = 1$ since the correct answer could be guessed or the respondent could be inattentive, producing $W_{pi2} + W_{pi3} = 0$. The log-likelihood function $l_{i2}^{\text{EM}}$ takes the form of the log-likelihood for multinomial data with one trial and with the regressor-based probabilities $\boldsymbol{Z}_p^\intercal\boldsymbol{c}_i$, $\boldsymbol{Z}_p^\intercal\boldsymbol{d}_i - \boldsymbol{Z}_p^\intercal\boldsymbol{c}_i$, and $1 - \boldsymbol{Z}_p^\intercal\boldsymbol{d}_i$.

The \gls{em} algorithm estimates item parameters in two steps -- expectation and maximization. These two steps are repeated until the convergence criterion is met, such as until the change in log-likelihood is lower than a predefined value. Since the \gls{em} algorithm is designed to obtain maximum likelihood estimates, their asymptotic properties are the same as described in Appendix~\ref{app:asymptotics:ml}. 

\paragraph{Expectation. }
At the E-step, conditionally on the item responses $Y_{pi}$ and the current parameter estimate $(\widehat{\boldsymbol{b}}_i, \widehat{\boldsymbol{c}}_i, \widehat{\boldsymbol{d}}_i)$, the estimates of latent variables are calculated as their expected values:
\begin{align}\label{eq:em:expectation}
    \begin{aligned}
        \widehat{W}_{pi1} =& \ \frac{\boldsymbol{Z}_p^\intercal\widehat{\boldsymbol{c}}_i Y_{pi}}{\boldsymbol{Z}_p^\intercal\widehat{\boldsymbol{c}}_i + (\boldsymbol{Z}_p^\intercal\widehat{\boldsymbol{d}}_i - \boldsymbol{Z}_p^\intercal\widehat{\boldsymbol{c}}_i) \widehat{\phi}_{pi}}, 
        \ \ \  &\widehat{W}_{pi2} =& \ Y_{pi} - \widehat{W}_{pi1}, \\
        \widehat{W}_{pi4} =& \ \frac{\left(1 - \boldsymbol{Z}_p^\intercal\widehat{\boldsymbol{d}}_i\right) \left(1 - Y_{pi}\right)}{1 - \boldsymbol{Z}_p^\intercal\widehat{\boldsymbol{c}}_i - (\boldsymbol{Z}_p^\intercal\widehat{\boldsymbol{d}}_i - \boldsymbol{Z}_p^\intercal\widehat{\boldsymbol{c}}_i) \widehat{\phi}_{pi}}, 
        \ \ \ &\widehat{W}_{pi3} =& \ 1 - Y_{pi} - \widehat{W}_{pi4}.
    \end{aligned}
\end{align}

\paragraph{Maximization. }
At the M-step, conditionally on the current estimates of the latent variables $\widehat{W}_{pi2}$ and $\widehat{W}_{pi3}$, the estimates of parameters $\boldsymbol{b}_i$ maximize the log-likelihood function $l_{i1}^{\text{EM}}$. The estimates $\widehat{\boldsymbol{c}}_i$ and $\widehat{\boldsymbol{d}}_i$ are given by a maximisation of the log-likelihood function $l_{i2}^{\text{EM}}$ conditionally on~current estimates of the latent variables $\widehat{W}_{pi1}$, $\widehat{W}_{pi2}$, $\widehat{W}_{pi3}$, and $\widehat{W}_{pi4}$.

The \gls{em} algorithm is designed to gain the \gls{ml} estimates of the item parameters, so estimates have the same asymptotic properties as described above. 

Additionally, it might be of practical interest that the \gls{em} algorithm provides estimates of latent variables $W_{pi1}$, $W_{pi2}$, $W_{pi3}$, and $W_{pi4}$. Their mean values over all items may be interpreted in an educational context as follows: The $\overline{W}_{p1}$ as (undesired) "inclination to guess"; $\overline{W}_{p2}$ as a (desired) probability of "knowing correct answers when correctly answering"; $\overline{W}_{p3}$ as a (desired) probability of "not knowing correct answers when incorrectly answering"; and $\overline{W}_{p4}$ as (undesired) "inclination to slip/inattention". In the context of psychological or health-related measurements, the mean values may be interpreted as follows: The $\overline{W}_{p1}$ as the (undesired) "inclination to simulate"; $\overline{W}_{p2}$ as the (desired) probability of "endorsing while experiencing described situations"; $\overline{W}_{p3}$ as the (desired) probability of "not endorsing while not experiencing described situations"; and $\overline{W}_{p4}$ as the (undesired) "inclination to dissimulate". 


\subsubsection{Parametrized link function}

In our setting, the covariate-specific \gls{4pl} model \eqref{eq:4PL:gen} can be viewed as a generalized linear model with a known \gls{plf} 
\begin{align}\label{eq:plf}
    g(\mu_{pi}; \boldsymbol{c}_i, \boldsymbol{d}_i) = \log\left(\frac{\mu_{pi} - \boldsymbol{Z}_p^\intercal\boldsymbol{c}_i}{\boldsymbol{Z}_p^\intercal\boldsymbol{d}_i - \mu_{pi}}\right),
\end{align}
where the parameters $\boldsymbol{c}_i$ and $\boldsymbol{d}_i$ are unknown and may depend on regressors $\boldsymbol{Z}_p$. Subsequently, the mean function is determined by $\mu_{pi} = \pi_{pi}$ as given by \eqref{eq:4PL:gen} with a linear predictor $\boldsymbol{X}_p^\intercal\boldsymbol{b}_i$. If the asymptote parameters were known, the estimation could proceed analogously to generalized linear models \cite<see, e.g.,>{dobson2018introduction}, specifically the standard logistic regression model. However, since the asymptote parameters are unknown, an additional step to estimate them is required.

Keeping this setting in mind, this study proposes a new two-stage algorithm to estimate item parameters using the \gls{plf} \eqref{eq:plf}, which involves repeating two steps until the convergence criterion is fulfilled. Similarly to the \gls{em} algorithm, the \gls{plf}-based estimation method is designed to compute maximum likelihood estimates; their asymptotic properties align with those detailed in Appendix~\ref{app:asymptotics:ml}.  

\paragraph{Step one. }
First, conditionally on current estimates $\widehat{\boldsymbol{c}}_i$ and $\widehat{\boldsymbol{d}}_i$ of the \gls{plf}, the estimates of~parameters $\boldsymbol{b}_i$ maximise the following log-likelihood function:
\begin{align*}
    l_{i1}^{\text{PL}}(\boldsymbol{b}_i|\widehat{\boldsymbol{c}}_i, \widehat{\boldsymbol{d}}_i) = \sum_{p = 1}^n &\left\{Y_{pi} \log(\boldsymbol{Z}_p^\intercal\widehat{\boldsymbol{c}}_i + (\boldsymbol{Z}_p^\intercal\widehat{\boldsymbol{d}}_i - \boldsymbol{Z}_p^\intercal\widehat{\boldsymbol{c}}_i)\phi_{pi}) \right. \\
     & \ \ \left. +\ (1 - Y_{pi})\log(1 - \boldsymbol{Z}_p^\intercal\widehat{\boldsymbol{c}}_i - (\boldsymbol{Z}_p^\intercal\widehat{\boldsymbol{d}}_i - \boldsymbol{Z}_p^\intercal\widehat{\boldsymbol{c}}_i)\phi_{pi})\right\}.
\end{align*}
The log-likelihood function $l_{i1}^{\text{PL}}(\boldsymbol{b}_i|\widehat{\boldsymbol{c}}_i, \widehat{\boldsymbol{d}}_i)$ has a similar form to the log-likelihood function $l_i(\boldsymbol{b}_i, \boldsymbol{c}_i, \boldsymbol{d}_i)$ using the \gls{ml} method. However, the parameters $\boldsymbol{c}_i$ and $\boldsymbol{d}_i$ are here replaced by their current estimates, $\widehat{\boldsymbol{c}}_i$ and $\widehat{\boldsymbol{d}}_i$. 

\paragraph{Step two. }
Next, estimates $\widehat{\boldsymbol{c}}_i$ and $\widehat{\boldsymbol{d}}_i$ of the \gls{plf} \eqref{eq:plf} are calculated conditionally on the current estimates $\widehat{\boldsymbol{b}}_i$ as the arguments of the maxima of the following log-likelihood function
\begin{align*}
    l_{i2}^{\text{PL}}(\boldsymbol{c}_i, \boldsymbol{d}_i|\widehat{\boldsymbol{b}}_i) = \sum_{p = 1}^n &\left\{Y_{pi} \log(\boldsymbol{Z}_p^\intercal\boldsymbol{c}_i + (\boldsymbol{Z}_p^\intercal\boldsymbol{d}_i - \boldsymbol{Z}_p^\intercal\boldsymbol{c}_i)\widehat{\phi}_{pi})\right. \\
    &\left. \ \ +\ (1 - Y_{pi})\log(1 - \boldsymbol{Z}_p^\intercal\boldsymbol{c}_i - (\boldsymbol{Z}_p^\intercal\boldsymbol{d}_i - \boldsymbol{Z}_p^\intercal\boldsymbol{c}_i)\widehat{\phi}_{pi})\right\}.
\end{align*}
Again, the parameters $\boldsymbol{b}_i$ are replaced by their estimates $\widehat{\boldsymbol{b}}_i$, and $\phi_{pi}$ is thus replaced by $\widehat{\phi}_{pi}$.

In summary, the division into the two sets of parameters makes the algorithm based on \gls{plf} easy to implement in the \pkg{R} software and can take advantage of its existing functions. Because the algorithm is designed to produce the \gls{ml} estimates, their asymptotic properties are the same as described above. 


\subsection{Implementation and software}

For all analyses, software \pkg{R}, version 4.3.1 \cite{r2023} was used. The methods proposed here are implemented into the \pkg{difNLR} package version 1.5.0 \cite{hladka2020difnlr}, and some of them are available in the interactive application of the \pkg{ShinyItemAnalysis} package \cite{martinkova2023computational, martinkova2018shinyitemanalysis}, version 1.5.3, see Figure (screenshot) in the electronic Supplementary Material. The \gls{nls} method was implemented using the base \Rcode{nls()} function and the \Rcode{"port"} algorithm \cite{port}. The sandwich estimator \eqref{app:eq:nls:sandwich} of the asymptotic covariance matrix was computed using the \pkg{calculus} package \cite{guidotti2022calculus}. The \gls{ml} estimation was performed with the base \Rcode{optim()} function and the \Rcode{"L-BFGS-B"} algorithm \cite{byrd1995limited}. The \gls{em} algorithm implements directly \eqref{eq:em:expectation} in the expectation step using the base \Rcode{glm()} function and the \Rcode{multinom()} function from the \pkg{nnet} package \cite{venables2002modern} in the maximization step. Next, step one of the newly proposed algorithm based on \gls{plf} is implemented with the base \Rcode{glm()} function with the modified logit link, which includes asymptote parameters. The asymptote parameters are estimated in step two using the base \Rcode{optim()} function. The maximum number of iterations was 2,000 for all four methods, and the convergence criterion was set to $10^{-6}$ when possible. 

\paragraph{Initial values. }
Starting values for item parameters were calculated as follows: The respondents were divided into three groups based upon tertiles of the matching criterion $\theta_p$. Next, the asymptote parameters were estimated: $c$ was computed as an empirical probability for those whose matching criterion was smaller than its average value in the first group defined by tertiles. The asymptote $d$ was calculated as an empirical probability of those whose matching criterion was greater than its average value in the last group defined by tertiles. The slope parameter $b_1$ was estimated as a difference between the mean empirical probabilities of the last and the first group multiplied by 4. This difference is sometimes called the upper-lower index. Finally, the intercept $b_0$ was calculated as follows: First, a center point between the asymptotes was computed, and then we looked for the level of the matching criterion that would have corresponded to this empirical probability. Additionally, smoothing and corrections for the variability of the matching criterion were applied.


\section{Simulation study}\label{sec:simulation}

A simulation study was performed to compare various procedures to estimate parameters in the generalized logistic regression model, including the \gls{nls}, the \gls{ml} method, the \gls{em} algorithm, and the newly proposed algorithm based on \gls{plf}. Two models were considered -- the simple \gls{4pl} model~\eqref{eq:4PL:simple} and the group-specific \gls{4pl} model~\eqref{eq:4PL:dif}.  


\subsection{Simulation design}

\paragraph{Data generation. }
To generate data, ten sets with different combinations of item parameters were considered. The item parameters were chosen to correspond to common values: Parameters $b_0$, $b_2$, and $b_3$ were generated from the standard normal distribution, parameter $b_1$ was generated from a normal distribution with a mean value equal to 2.5, and a standard deviation of 0.5. Parameter $c$ was generated from uniform distribution $\mathcal{U}(0.05, 0.30)$ for both groups. Parameter $d$ was generated from uniform distribution $\mathcal{U}(0.7, 0.95)$ for both groups. In the case of the simple \gls{4pl} model \eqref{eq:4PL:simple}, only parameters $b_0$, $b_1$, $c$, and $d$ were considered. Next, the matching criterion $\theta_p$ was generated from the standard normal distribution for all respondents. Since item parameters are usually estimated item by item in generalized logistic models, we focused on items separately, and only one item was generated for each scenario. We used a generated variable as the matching criterion instead of the total score. Binary responses were generated from the Bernoulli distribution with the calculated probabilities based upon the chosen \gls{4pl} model, true parameters, and the matching criterion variable. The sample size was set to $n =$ 500; 1,000; 2,500; and 5,000, i.e., 250; 500; 1,250; and 2,500 per group in the case of the group-specific \gls{4pl} model \eqref{eq:4PL:dif}. Each scenario was replicated 1,000 times. 

\paragraph{Simulation evaluation. }
To compare estimation methods, we first computed mean and median numbers of iteration runs and the convergence status of the methods, meaning the percentage of converged simulation runs; the percentage of runs that crashed (caused an error when fitting, e.g., due to singularities); and the percentage of those which reached the maximum number of iterations without convergence. Next, we selected only those simulation runs for which all four estimation methods converged successfully. We computed the mean parameter estimates and parametric confidence intervals, i.e., average intervals found for estimated standard errors derived for the respective algorithm. When confidence intervals for asymptote parameters exceeded their boundaries of 0 or 1, confidence intervals were truncated at the boundary value. The proportion of confidence intervals covering the true parameter value was calculated. Subsequently, the mean bias (i.e., the mean difference between estimates and true values) and \gls{rmse} (i.e., the square root of the average of squared errors) were calculated with respect to sample size. Finally, for a deeper insight into \gls{ml}-based methods (i.e., traditionally implemented \gls{ml}, the \gls{em} algorithm, and the algorithm based on \gls{plf}), we compared log-likelihoods for these three methods to those based on true values of parameters. 

\subsection{Simulation results}

\paragraph{Convergence status. }
All four methods had low percentages of simulation runs that crashed for all sample sizes in the simple \gls{4pl} model \eqref{eq:4PL:simple}. Still, the rate was mildly increased in the group-specific \gls{4pl} model \eqref{eq:4PL:dif} for the \gls{nls} method (6.72\%) and for the algorithm based on \gls{plf} (9.22\%) when $n = 500$. With the increasing sample size, convergence issues disappeared. The \gls{em} algorithm struggled to converge in a predefined number of iterations, especially for small sample sizes in both models. Additionally, the method based on \gls{plf} reached the maximum limit of 2,000 iterations only in a small percentage of simulation runs when smaller sample sizes were considered (Table \ref{tab:convergence}). 

\paragraph{Number of iterations. }
Furthermore, the methods differed in the number of iterations needed until the estimation process successfully ended. The \gls{em} algorithm yielded the largest mean and median numbers of iterations, which were somehow overestimated by simulation runs that did not finish without convergence (i.e., the maximum limit of 2,000 iterations was reached). The fewest iterations were needed for the \gls{nls} method. As expected, all the procedures required fewer simulation runs when the simple \gls{4pl} model \eqref{eq:4PL:simple} was considered than in the group-specific \gls{4pl} model \eqref{eq:4PL:dif}. Beyond this, the number of iterations was decreasing with the increasing sample size in both models for all the methods (Table \ref{tab:convergence}).

\vspace{1em}

In both models, some estimation procedures produced non-meaningful estimates of parameters $b_0$--$b_3$ (absolute value over 100) despite successful convergence. Such simulations affected mean values significantly, so they were removed from a computation of the mean estimates and their confidence intervals for all four estimation methods. Incidence was similar for all methods (Table~\ref{tab:convergence}). Such estimates might be obtained due to insufficient sample size or starting values far from the global maximizer. 

\begin{table}[ht]
\centering
\caption{Convergence status, proportion of suspicious simulation runs, and the number of iterations for the four estimation methods} 
\label{tab:convergence}
\begin{threeparttable}
\begin{tabular}{@{}p{\textwidth}@{}}
\centering
\resizebox{\textwidth}{!}{%
\begin{tabular}{lrrrrrrrrrrrr}
  \toprule 
 \multirow{3}{*}{Method} & \multicolumn{6}{c}{Simple \gls{4pl} model \eqref{eq:4PL:simple}} & \multicolumn{6}{c}{Group-specific \gls{4pl} model \eqref{eq:4PL:dif}} \\ \cmidrule(lr){2-7} \cmidrule(lr){8-13}
 & \multicolumn{4}{c}{Convergence status [\%]} & \multicolumn{2}{c}{Number of iterations} & \multicolumn{4}{c}{Convergence status [\%]} & \multicolumn{2}{c}{Number of iterations} \\ \cmidrule(lr){2-5} \cmidrule(lr){6-7} \cmidrule(lr){8-11} \cmidrule(lr){12-13}  
 & Conv. & Crash. & DNF & Susp. & Mean & Median & Conv. & Crash. & DNF & Susp. & Mean & Median \\ 
  \midrule \multicolumn{12}{c}{$n$ = 500} \\ \midrule 
NLS & 99.48 & 0.52 & 0.00 & 0.06 & 10.18 & 8.00 & 93.28 & 6.72 & 0.00 & 0.50 & 17.27 & 14.00 \\ 
  MLE & 99.84 & 0.16 & 0.00 & 0.06 & 23.84 & 23.00 & 98.87 & 1.13 & 0.00 & 0.45 & 123.75 & 104.00 \\ 
  EM & 93.70 & 0.00 & 6.30 & 0.03 & 347.14 & 152.00 & 93.38 & 0.07 & 6.55 & 0.50 & 452.89 & 211.00 \\ 
  PLF & 98.34 & 1.02 & 0.64 & 0.03 & 144.31 & 18.00 & 89.84 & 9.22 & 0.94 & 0.47 & 248.06 & 66.00 \\ 
   \midrule \multicolumn{12}{c}{$n$ = 1,000} \\ \midrule 
NLS & 99.93 & 0.07 & 0.00 & 0.00 & 7.66 & 7.00 & 97.90 & 2.10 & 0.00 & 0.15 & 12.50 & 10.00 \\ 
  MLE & 99.89 & 0.11 & 0.00 & 0.00 & 22.24 & 22.00 & 99.99 & 0.01 & 0.00 & 0.06 & 106.85 & 99.00 \\ 
  EM & 95.76 & 0.00 & 4.24 & 0.00 & 295.53 & 149.00 & 93.50 & 0.00 & 6.50 & 0.10 & 447.87 & 201.00 \\ 
  PLF & 99.71 & 0.11 & 0.18 & 0.00 & 97.73 & 14.00 & 97.84 & 1.96 & 0.20 & 0.10 & 172.43 & 47.00 \\ 
   \midrule \multicolumn{12}{c}{$n$ = 2,500} \\ \midrule 
NLS & 99.98 & 0.02 & 0.00 & 0.00 & 5.97 & 6.00 & 99.39 & 0.61 & 0.00 & 0.01 & 8.17 & 7.00 \\ 
  MLE & 99.94 & 0.06 & 0.00 & 0.00 & 20.98 & 20.00 & 100.00 & 0.00 & 0.00 & 0.00 & 94.19 & 91.00 \\ 
  EM & 96.23 & 0.00 & 3.77 & 0.00 & 254.47 & 134.00 & 94.55 & 0.00 & 5.45 & 0.01 & 356.17 & 161.00 \\ 
  PLF & 99.98 & 0.00 & 0.02 & 0.00 & 51.77 & 9.00 & 99.87 & 0.12 & 0.01 & 0.01 & 95.36 & 25.00 \\ 
   \midrule \multicolumn{12}{c}{$n$ = 5,000} \\ \midrule 
NLS & 100.00 & 0.00 & 0.00 & 0.00 & 5.26 & 5.00 & 99.97 & 0.03 & 0.00 & 0.00 & 6.57 & 6.00 \\ 
  MLE & 99.95 & 0.05 & 0.00 & 0.00 & 20.50 & 20.00 & 100.00 & 0.00 & 0.00 & 0.00 & 90.21 & 89.00 \\ 
  EM & 97.64 & 0.00 & 2.36 & 0.00 & 223.45 & 127.00 & 95.90 & 0.00 & 4.10 & 0.01 & 308.32 & 141.00 \\ 
  PLF & 100.00 & 0.00 & 0.00 & 0.00 & 24.23 & 8.00 & 99.98 & 0.02 & 0.00 & 0.02 & 50.03 & 11.00 \\ 
   \bottomrule 
\end{tabular}%
}
\end{tabular}
\tablenotet{Conv. = converged, Crash. = crashed, DNF = did not finish, Susp. = suspicious. }
\end{threeparttable}
\end{table}

\paragraph{Parameter estimates. }
In the simple \gls{4pl} model \eqref{eq:4PL:simple}, the \gls{plf}-based algorithm gained the most precise estimates of parameters $b_0$ and $b_1$ (in the sense of bias and \gls{rmse}) when smaller sample sizes were considered ($n = 500$ or $n = 1,000$). Additionally, the \gls{nls} method yielded slightly more biased estimates in these scenarios. The precision of the estimation improved for both parameters when the sample size increased in all four methods, whereas differences between estimation procedures narrowed. The accuracy of the estimates of the asymptote parameters $c$ and $d$ was similar for all four methods. The \gls{nls} method yielded the least biased estimates, while the \gls{plf}-based algorithm produced the lowest \gls{rmse}. However, the differences between estimation approaches were minor. The proportion of confidence intervals covering true values of item parameters was high for all four methods (Table~\ref{tab:simulation:pars_simple}). Slightly higher coverage for the \gls{nls} method was caused by somewhat larger confidence intervals. 

\begin{table}[ht]
\centering
\caption{Bias, \gls{rmse}, and confidence interval (CI) coverage by four estimation methods using the simple model \eqref{eq:4PL:simple} with respect to sample size} 
\label{tab:simulation:pars_simple}
\begin{adjustbox}{width=1\textwidth}
\begin{tabular}{lrrrrrrrrc}
  \toprule 
 \multirow{2}{*}{Method} & \multicolumn{4}{c}{Bias} & \multicolumn{4}{c}{\gls{rmse}} & \multirow{2}{*}{\shortstack[c]{\\[0.3ex] CI \\[0.4ex] coverage [\%]}} \\ 
                               \cmidrule(lr){2-5} \cmidrule(lr){6-9}  
 & \multicolumn{1}{c}{500} & \multicolumn{1}{c}{1,000} & \multicolumn{1}{c}{2,500} & \multicolumn{1}{c}{5,000} & \multicolumn{1}{c}{500} & \multicolumn{1}{c}{1,000} & \multicolumn{1}{c}{2,500} & \multicolumn{1}{c}{5,000} &  \\ 
  \midrule \multicolumn{10}{l}{$b_0$} \\ 
\hspace{1em} NLS & $0.044$ & $0.012$ & $0.003$ & $0.003$ & $1.028$ & $0.399$ & $0.208$ & $0.143$ & $96.27$ \\ 
  \hspace{1em} MLE & $0.045$ & $0.016$ & $0.004$ & $0.004$ & $0.902$ & $0.387$ & $0.204$ & $0.141$ & $95.89$ \\ 
  \hspace{1em} EM & $0.035$ & $0.015$ & $0.003$ & $0.003$ & $0.839$ & $0.388$ & $0.204$ & $0.141$ & $95.75$ \\ 
  \hspace{1em} PLF & $0.016$ & $0.010$ & $0.006$ & $0.006$ & $0.514$ & $0.322$ & $0.196$ & $0.144$ & $94.95$ \\ 
   \midrule \multicolumn{10}{l}{$b_1$} \\ 
\hspace{1em} NLS & $-0.645$ & $-0.217$ & $-0.070$ & $-0.037$ & $2.876$ & $0.990$ & $0.446$ & $0.300$ & $95.44$ \\ 
  \hspace{1em} MLE & $-0.507$ & $-0.183$ & $-0.059$ & $-0.031$ & $2.253$ & $0.932$ & $0.433$ & $0.292$ & $95.60$ \\ 
  \hspace{1em} EM & $-0.470$ & $-0.174$ & $-0.055$ & $-0.026$ & $2.194$ & $0.922$ & $0.430$ & $0.291$ & $95.47$ \\ 
  \hspace{1em} PLF & $-0.159$ & $-0.033$ & $0.024$ & $0.026$ & $1.011$ & $0.628$ & $0.377$ & $0.278$ & $95.19$ \\ 
   \midrule \multicolumn{10}{l}{$c$} \\ 
\hspace{1em} NLS & $0.003$ & $0.003$ & $0.001$ & $0.000$ & $0.061$ & $0.044$ & $0.027$ & $0.019$ & $94.62$ \\ 
  \hspace{1em} MLE & $0.006$ & $0.004$ & $0.001$ & $0.001$ & $0.063$ & $0.044$ & $0.027$ & $0.019$ & $94.34$ \\ 
  \hspace{1em} EM & $0.006$ & $0.004$ & $0.002$ & $0.001$ & $0.062$ & $0.043$ & $0.027$ & $0.019$ & $94.27$ \\ 
  \hspace{1em} PLF & $0.012$ & $0.009$ & $0.005$ & $0.004$ & $0.060$ & $0.042$ & $0.027$ & $0.019$ & $95.16$ \\ 
   \midrule \multicolumn{10}{l}{$d$} \\ 
\hspace{1em} NLS & $-0.001$ & $-0.001$ & $0.000$ & $0.000$ & $0.065$ & $0.049$ & $0.032$ & $0.023$ & $94.28$ \\ 
  \hspace{1em} MLE & $-0.003$ & $-0.002$ & $-0.000$ & $0.000$ & $0.065$ & $0.049$ & $0.031$ & $0.022$ & $93.88$ \\ 
  \hspace{1em} EM & $-0.003$ & $-0.001$ & $-0.000$ & $0.000$ & $0.065$ & $0.049$ & $0.031$ & $0.022$ & $93.77$ \\ 
  \hspace{1em} PLF & $-0.009$ & $-0.006$ & $-0.003$ & $-0.002$ & $0.062$ & $0.047$ & $0.030$ & $0.022$ & $94.12$ \\ 
   \bottomrule 
\end{tabular}
\end{adjustbox}
\end{table}

In the group-specific \gls{4pl} model \eqref{eq:4PL:dif}, the \gls{plf}-based algorithm yielded the most precise estimates of parameters $b_0$--$b_3$ in the sense of \gls{rmse}, especially for the smaller sample sizes. On the other hand, the \gls{nls} method produced the largest \gls{rmse} in such scenarios. Computed bias was similar for all four methods. Similar to the simple \gls{4pl} model \eqref{eq:4PL:simple}, the differences in the precision of the parameter estimates were narrowed with the increasing sample size, and all four estimation approaches gave estimates close to the true values of the item parameters. The estimates of the asymptote parameters $c$, $c_{\text{DIF}}$, $d$, and $d_{\text{DIF}}$ were similar for all four methods. The \gls{em} algorithm provided slightly less biased mean estimates of the asymptote parameters, while the \gls{plf}-based algorithm produced slightly smaller \gls{rmse}. The proportion of confidence intervals covering true values of item parameters was high and similar for all four methods (Table~\ref{tab:simulation:pars_DIF}). Different lengths of computed intervals caused differences in coverage of true parameters between the estimation methods. 

\begin{table}[h!]
\centering
\caption{Bias, \gls{rmse}, and confidence interval (CI) coverage by four estimation methods using the group-specific model \eqref{eq:4PL:dif} with respect to sample size} 
\label{tab:simulation:pars_DIF}
\begin{adjustbox}{width=1\textwidth}
\begin{tabular}{lrrrrrrrrc}
  \toprule 
 \multirow{2}{*}{Method} & \multicolumn{4}{c}{Bias} & \multicolumn{4}{c}{\gls{rmse}} & \multirow{2}{*}{\shortstack[c]{\\[0.3ex] CI \\[0.4ex] coverage [\%]}} \\ 
                               \cmidrule(lr){2-5} \cmidrule(lr){6-9}  
 & \multicolumn{1}{c}{500} & \multicolumn{1}{c}{1,000} & \multicolumn{1}{c}{2,500} & \multicolumn{1}{c}{5,000} & \multicolumn{1}{c}{500} & \multicolumn{1}{c}{1,000} & \multicolumn{1}{c}{2,500} & \multicolumn{1}{c}{5,000} &  \\ 
  \midrule \multicolumn{10}{l}{$b_0$} \\ 
\hspace{1em} NLS & $-0.025$ & $0.008$ & $0.007$ & $0.003$ & $2.230$ & $1.259$ & $0.351$ & $0.204$ & $96.63$ \\ 
  \hspace{1em} MLE & $0.002$ & $0.023$ & $0.009$ & $0.004$ & $1.777$ & $1.105$ & $0.341$ & $0.200$ & $96.07$ \\ 
  \hspace{1em} EM & $0.023$ & $0.026$ & $0.008$ & $0.005$ & $1.770$ & $1.086$ & $0.337$ & $0.200$ & $95.40$ \\ 
  \hspace{1em} PLF & $0.009$ & $0.006$ & $0.010$ & $0.011$ & $0.846$ & $0.485$ & $0.285$ & $0.232$ & $94.71$ \\ 
   \midrule \multicolumn{10}{l}{$b_1$} \\ 
\hspace{1em} NLS & $-1.400$ & $-0.640$ & $-0.156$ & $-0.072$ & $6.027$ & $3.041$ & $0.896$ & $0.445$ & $94.64$ \\ 
  \hspace{1em} MLE & $-1.013$ & $-0.506$ & $-0.122$ & $-0.059$ & $4.745$ & $2.463$ & $0.768$ & $0.431$ & $95.37$ \\ 
  \hspace{1em} EM & $-0.811$ & $-0.475$ & $-0.116$ & $-0.053$ & $4.289$ & $2.371$ & $0.768$ & $0.427$ & $94.12$ \\ 
  \hspace{1em} PLF & $-0.214$ & $-0.120$ & $0.010$ & $0.036$ & $1.510$ & $1.000$ & $0.541$ & $0.390$ & $93.67$ \\ 
   \midrule \multicolumn{10}{l}{$b_2$} \\ 
\hspace{1em} NLS & $0.082$ & $0.016$ & $-0.001$ & $0.002$ & $3.005$ & $1.528$ & $0.477$ & $0.289$ & $96.91$ \\ 
  \hspace{1em} MLE & $0.006$ & $-0.012$ & $-0.007$ & $-0.001$ & $2.467$ & $1.357$ & $0.463$ & $0.284$ & $96.34$ \\ 
  \hspace{1em} EM & $-0.022$ & $-0.011$ & $-0.006$ & $0.000$ & $2.270$ & $1.277$ & $0.461$ & $0.283$ & $95.59$ \\ 
  \hspace{1em} PLF & $0.024$ & $0.005$ & $-0.012$ & $-0.013$ & $1.319$ & $0.763$ & $0.401$ & $0.309$ & $94.53$ \\ 
   \midrule \multicolumn{10}{l}{$b_3$} \\ 
\hspace{1em} NLS & $0.047$ & $0.031$ & $-0.000$ & $-0.003$ & $8.105$ & $4.068$ & $1.175$ & $0.649$ & $97.73$ \\ 
  \hspace{1em} MLE & $-0.113$ & $0.023$ & $-0.003$ & $-0.004$ & $6.367$ & $3.366$ & $1.040$ & $0.630$ & $97.46$ \\ 
  \hspace{1em} EM & $0.098$ & $0.061$ & $0.003$ & $0.001$ & $5.889$ & $3.046$ & $1.040$ & $0.625$ & $95.93$ \\ 
  \hspace{1em} PLF & $0.084$ & $0.101$ & $0.065$ & $0.050$ & $2.329$ & $1.532$ & $0.800$ & $0.584$ & $95.06$ \\ 
   \midrule \multicolumn{10}{l}{$c$} \\ 
\hspace{1em} NLS & $0.015$ & $0.005$ & $0.003$ & $0.001$ & $0.087$ & $0.062$ & $0.039$ & $0.027$ & $94.25$ \\ 
  \hspace{1em} MLE & $0.020$ & $0.008$ & $0.004$ & $0.001$ & $0.091$ & $0.063$ & $0.040$ & $0.027$ & $94.23$ \\ 
  \hspace{1em} EM & $0.024$ & $0.008$ & $0.004$ & $0.001$ & $0.091$ & $0.063$ & $0.039$ & $0.027$ & $93.33$ \\ 
  \hspace{1em} PLF & $0.029$ & $0.015$ & $0.009$ & $0.006$ & $0.089$ & $0.061$ & $0.039$ & $0.027$ & $94.51$ \\ 
   \midrule \multicolumn{10}{l}{$c_{\text{DIF}}$} \\ 
\hspace{1em} NLS & $-0.007$ & $-0.002$ & $-0.001$ & $-0.001$ & $0.113$ & $0.083$ & $0.053$ & $0.037$ & $95.19$ \\ 
  \hspace{1em} MLE & $-0.007$ & $-0.002$ & $-0.001$ & $-0.001$ & $0.116$ & $0.085$ & $0.053$ & $0.037$ & $95.13$ \\ 
  \hspace{1em} EM & $-0.004$ & $-0.002$ & $-0.001$ & $-0.001$ & $0.117$ & $0.084$ & $0.052$ & $0.037$ & $93.63$ \\ 
  \hspace{1em} PLF & $-0.005$ & $-0.000$ & $0.001$ & $0.001$ & $0.110$ & $0.080$ & $0.052$ & $0.037$ & $94.58$ \\ 
   \midrule \multicolumn{10}{l}{$d$} \\ 
\hspace{1em} NLS & $-0.010$ & $-0.003$ & $-0.003$ & $-0.001$ & $0.084$ & $0.065$ & $0.044$ & $0.031$ & $93.72$ \\ 
  \hspace{1em} MLE & $-0.015$ & $-0.005$ & $-0.004$ & $-0.001$ & $0.087$ & $0.066$ & $0.045$ & $0.031$ & $93.46$ \\ 
  \hspace{1em} EM & $-0.018$ & $-0.005$ & $-0.003$ & $-0.001$ & $0.088$ & $0.065$ & $0.043$ & $0.031$ & $91.56$ \\ 
  \hspace{1em} PLF & $-0.022$ & $-0.012$ & $-0.008$ & $-0.005$ & $0.084$ & $0.063$ & $0.043$ & $0.031$ & $92.31$ \\ 
   \midrule \multicolumn{10}{l}{$d_{\text{DIF}}$} \\ 
\hspace{1em} NLS & $0.003$ & $0.002$ & $0.001$ & $0.000$ & $0.118$ & $0.091$ & $0.059$ & $0.043$ & $95.29$ \\ 
  \hspace{1em} MLE & $0.003$ & $0.002$ & $0.002$ & $0.000$ & $0.121$ & $0.092$ & $0.060$ & $0.043$ & $95.17$ \\ 
  \hspace{1em} EM & $-0.000$ & $0.001$ & $0.001$ & $0.000$ & $0.121$ & $0.091$ & $0.058$ & $0.042$ & $92.77$ \\ 
  \hspace{1em} PLF & $0.000$ & $-0.000$ & $-0.001$ & $-0.001$ & $0.114$ & $0.087$ & $0.058$ & $0.042$ & $93.57$ \\ 
   \bottomrule 
\end{tabular}
\end{adjustbox}
\end{table}

\paragraph{Log-likelihood comparison. }
In the simple \gls{4pl} model \eqref{eq:4PL:simple}, the algorithm based on \gls{plf} yielded log-likelihood values nearest to those computed based on true parameters in 91.31\% of cases, followed by the \gls{em} algorithm in 8.45\% and the directly implemented \gls{ml} method in 0.23\% of cases. There were similar differences between the three \gls{ml}-based methods in the group-specific \gls{4pl} model \eqref{eq:4PL:dif}. The algorithm based on \gls{plf} outperformed other likelihood-based estimation procedures in 87.56\% of cases, while the \gls{em} algorithm worked the best in 12.06\% and the \gls{ml} method in 0.38\%. 


\section{Real data examples}\label{sec:real-data}


\subsection{Data description}

We demonstrate the estimation procedures with an application to \gls{dif} detection on two real-data examples, which are available in the \pkg{ShinyItemAnalysis} \pkg{R} package and interactive application \cite{martinkova2018shinyitemanalysis, martinkova2023computational}, namely PROMIS anxiety scale\footnote{http://www.nihpromis.org} and a test measuring learning competence \cite{martinkova2020academic}. 


\subsubsection{Anxiety scale}

The Anxiety dataset consisted of responses to 29 Likert-type questions (1 = Never, 2 = Rarely, 3 = Sometimes, 4 = Often, and 5 = Always) from 766 respondents. Additionally, the dataset included information on the respondents' age, education, and gender (0 = Male and 1 = Female). Overall, there were 369 male participants and 397 female participants.  

For this work, item responses were dichotomized as follows: 0 was assigned to response Never (i.e., response $= 1$ on the original scale), while 1 was given to responses Rarely and more often (i.e., response $\geq 2$ on the original scale). 


\subsubsection{Learning competence}

The LearningToLearn dataset consisted of binary-coded responses from 782 subjects to (mostly) multiple-choice test consisting of 41 items within seven subscales. Each respondent was tested twice – the first time in the 6th grade and the second time in the 9th grade; responses from the 6th grade only were considered for this analysis. Among other variables, the dataset included information on the school track of respondents (basic school track = 0, academic school track = 1). Overall, 391 students attended basic school, and 391 pursued selective academic school. 


\subsection{Real data analysis design}

This work considered the simple \gls{4pl} model \eqref{eq:4PL:simple} and the group-specific \gls{4pl} model \eqref{eq:4PL:dif} with different constraints on asymptote parameters, yielding the \gls{3pl} models. In the Anxiety dataset, the lower asymptotes were set to zeros, i.e., $c_i = 0$ and $c_{i\text{DIF}} = 0$, since pretending (i.e., lower asymptote greater than 0) was not expected. On the other hand, in the LearningToLearn dataset, the upper asymptotes were set to ones, i.e., $d_i = 1$ and $d_{i\text{DIF}} = 0$, since slipping (i.e., upper asymptote lower than 1) was not expected. 

The matching criterion $\theta_p$ in the Anxiety dataset was the overall level of anxiety, which was calculated as a standardized sum of non-dichotomized item responses. Similarly, the standardized total test score gained in the 6th grade was used as the matching criterion $\theta_p$ in the LearningToLearn dataset. For the group-specific models, the grouping variable $G_p$ was defined by the gender of respondents for the Anxiety dataset and the school track for the LearningToLearn dataset. \gls{dif} detection was performed concerning these variables. 

The two newly proposed estimation methods were applied for the two datasets and models: the \gls{em} algorithm and the algorithm based on \gls{plf}. The same approach for computing starting values as in the simulation study was used to analyze both datasets. In the case of convergence issues, the initial values were re-calculated based on successfully converged estimates provided by other methods. 

In this study, item parameter estimates were computed and reported with confidence intervals. Next, the likelihood-ratio test \eqref{eq:lrt} was performed to compare the two nested models (simple and group-specific) to identify the \gls{dif} for all items and both novel estimation methods. Finally, for the \gls{em} algorithm, mean values of estimated latent variables over all items were computed. A significance level of 0.05 was used for all the tests.  


\subsection{Real data analysis results}


\subsubsection{Anxiety scale}

\paragraph{DIF detection. }
Using the likelihood-ratio test, the simple \gls{4pl} model \eqref{eq:4PL:simple} with constraints on lower asymptotes was rejected for items R6 (\highlight{"I was concerned about my mental health"}; $p$-value = 0.001 considering either estimation algorithm), R7 (\highlight{"I felt upset"}; $p$-value = 0.038), R10 (\highlight{"I had sudden feelings of panic"}; $p$-value = 0.031), R21 (\highlight{"I had twitching or trembling muscles"}; $p$-value = 0.035), and R29 (\highlight{"I had difficulty calming down"}; $p$-value = 0.04) by using either of the two newly proposed estimation algorithms (i.e., these items functioned differently). In these items, the less restrictive group-specific model \eqref{eq:4PL:dif} was preferred, allowing for different intercepts, slopes, and upper asymptotes for the two groups (i.e., these items functioned differently). 

\begin{figure}[h!]
    \centering
    \includegraphics[width=0.5\textwidth]{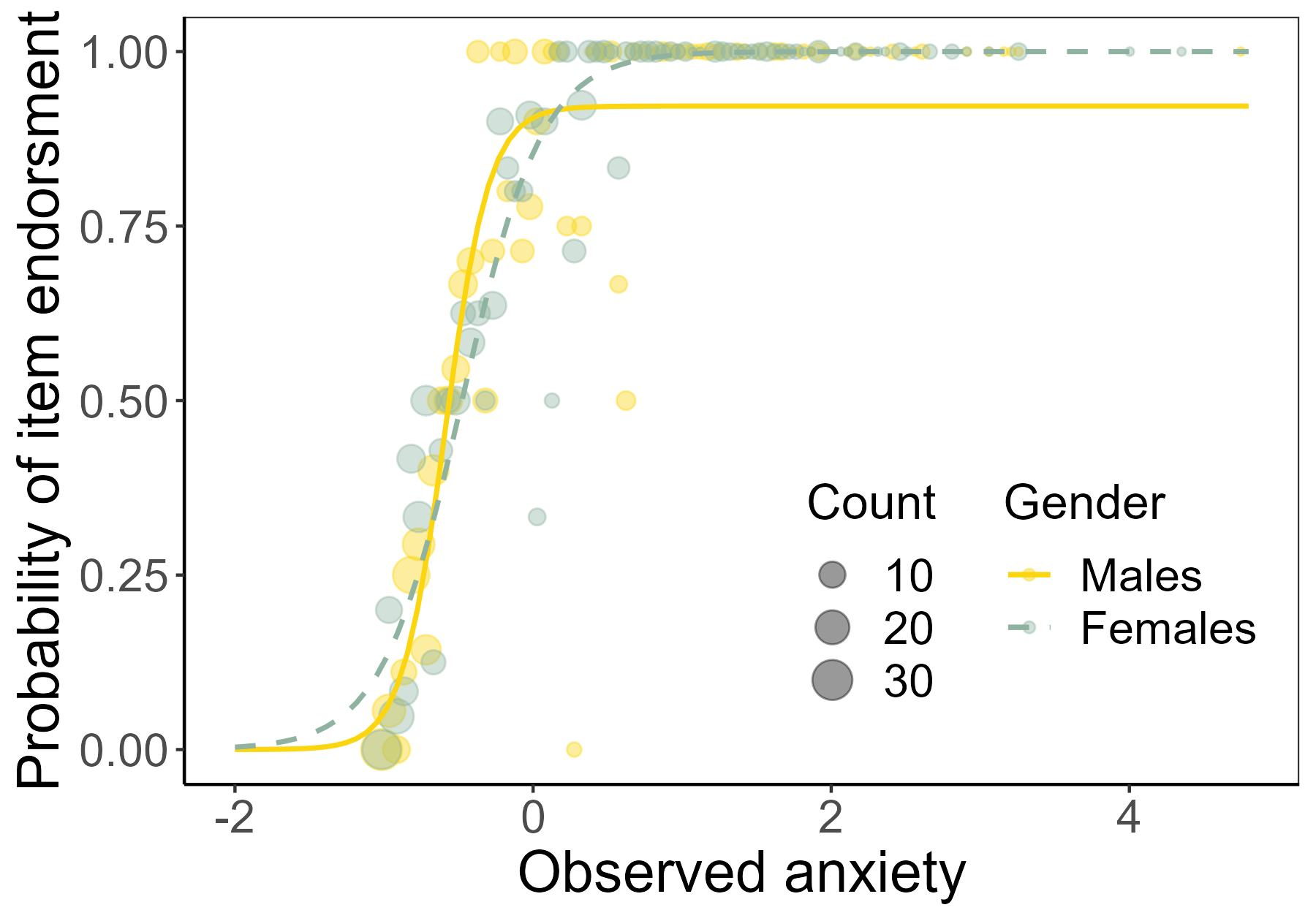}
   \caption{Estimated item characteristic curves of item R7 of the Anxiety dataset for the group-specific \gls{4pl} model \eqref{eq:4PL:dif} with constraints on lower asymptotes}\label{fig:Anxiety_icc_R7}
\end{figure}

We now take a closer look at \gls{dif} item R7, for which the confidence interval of estimated parameter $d_{\text{DIF}}$ (i.e., the difference in upper asymptote between groups of female and male respondents) did not cover 0 (Table~\ref{app:tab:anxiety:pars_dif}). According to the model, the male respondents with high anxiety levels seemed not to admit feeling upset with a probability of $1 - 0.92 = 0.08$, while there was no dissimulation rate for female respondents (Figure~\ref{fig:Anxiety_icc_R7}). 

\paragraph{Latent variables estimates by EM algorithm. }
Using the group-specific model~\eqref{eq:4PL:dif}, male respondent 272 with an overall level of anxiety equal to 0.62 had the highest "inclination to dissimulate," equal to 0.20, meaning they would have dissimulated almost six items out of the 29-item Anxiety dataset. On the other hand, female respondent 264 with the same overall anxiety level had a probability equal to 0.06, which would correspond to the dissimulation of less than two items. 


\subsubsection{Learning competence}

\paragraph{DIF detection. }
Using the likelihood-ratio test, the simple \gls{4pl} model \eqref{eq:4PL:simple} with constraints on upper asymptotes was rejected for items 1A ($p$-value = 0.006 considering either estimation algorithm), 1D ($p$-value = 0.009), 6F ($p$-value = 0.043), 6H ($p$-value = 0.032), and 7F ($p$-value = 0.049) by using either of the two newly proposed estimation algorithms. In these items, the less restrictive group-specific model \eqref{eq:4PL:dif} was preferred, allowing for different intercepts, slopes, and lower asymptotes for the two groups. 

\begin{figure}[h!]
    \centering
    \begin{subfigure}[t]{0.495\textwidth}
        \includegraphics[width=\textwidth]{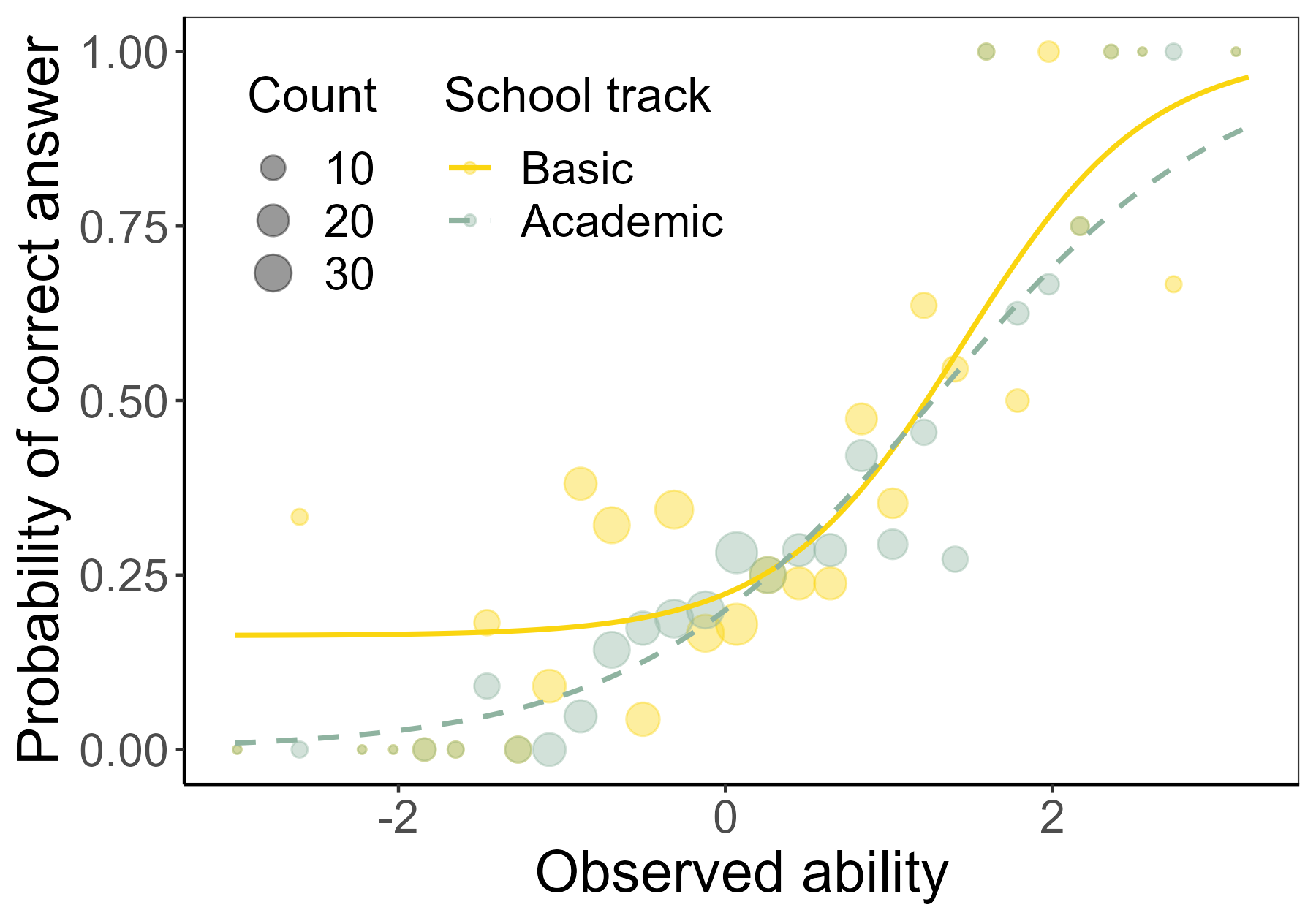}
        \caption{Item 6F}
        \label{fig:ltl_icc_6f}
    \end{subfigure}
    \begin{subfigure}[t]{0.495\textwidth}
        \includegraphics[width=\textwidth]{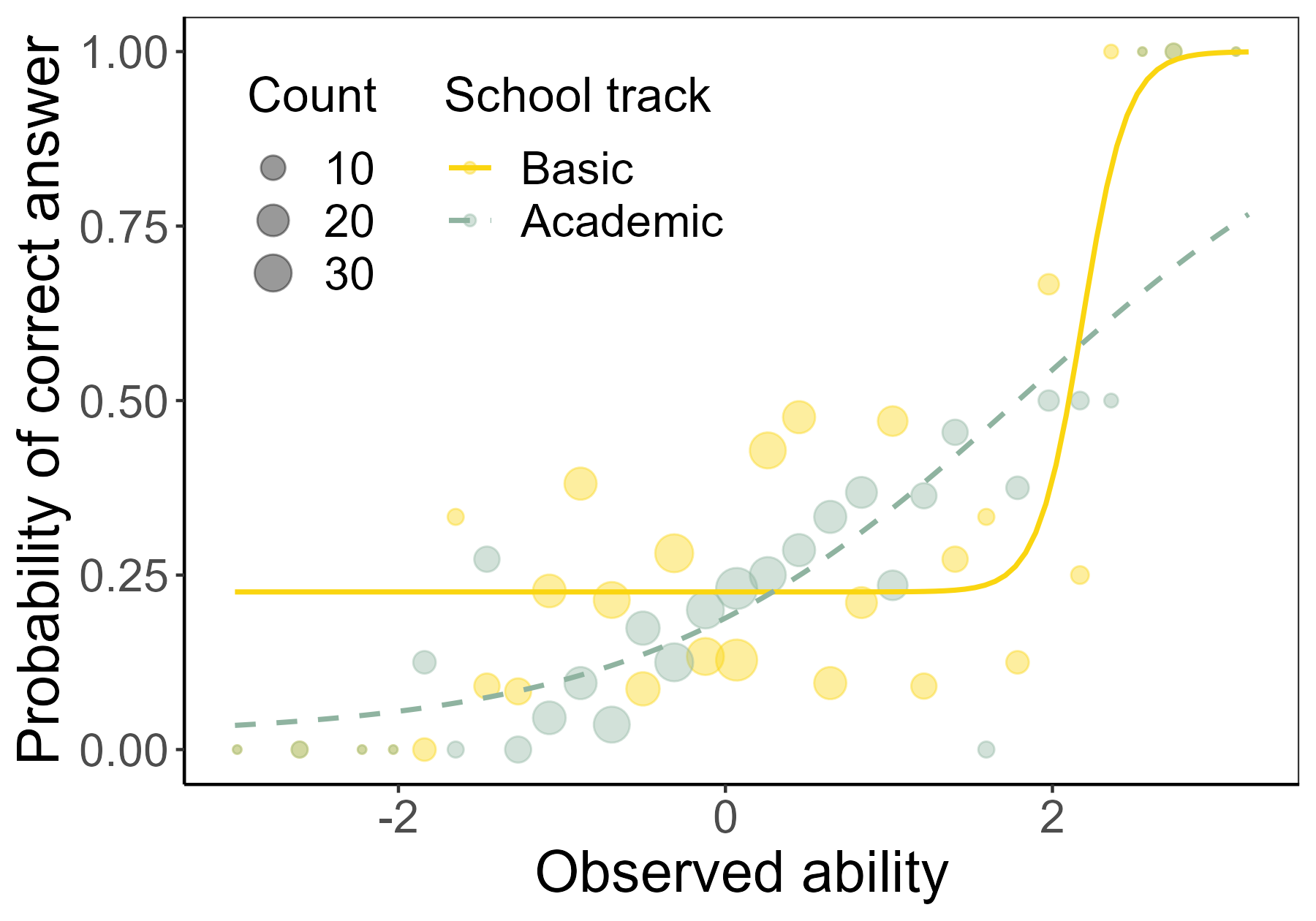}
        \caption{Item 6H}
        \label{fig:ltl_icc_6h}
    \end{subfigure}
   \caption{Estimated item characteristic curves of selected \gls{dif} items of the LearningToLearn dataset for the group-specific \gls{4pl} model \eqref{eq:4PL:dif} with constraints on upper asymptotes}\label{fig:ltl_icc}
\end{figure}

Items 6F and 6H were identified as functioning differently due to differences in lower asymptotes, i.e., confidence interval of estimated parameter $c_{\text{DIF}}$ (i.e., the difference in lower asymptotes between the two school tracks) did not cover 0 (Table~\ref{app:tab:ltl:pars_dif}). In both items, students from the basic school track tended to guess more often than students from the academic school track. In item 6F, the probability of guessing in the basic school track was 0.16, while in the academic school track, it was 0 (Figure~\ref{fig:ltl_icc_6f}).  In item 6H, the probability of guessing in the basic school track was 0.23, while in the academic school track, it was 0.02 (Figure~\ref{fig:ltl_icc_6h}). 

Both items were related to "solving tasks with invented mathematical operators which are conditionally defined depending on the value of the digits they connect" \cite{martinkova2020academic}. The original study suggested that students from academic schools might have been trying to solve these difficult items more often, while students from the basic school track might have been guessing more often. 

\paragraph{Latent variables estimates by EM algorithm. }
Considering the group-specific model~\eqref{eq:4PL:dif}, respondent 486, who attended basic school with an overall level of learning competence of $-0.70$, has the highest "inclination to guess," equal to 0.38, meaning they would guess almost 16 items out of the 41-item test on learning competencies. On the other hand, respondent 386, who attended academic school and who had exactly the same level of learning competence, has a probability equal to 0.17, corresponding to the guessing of 7 items out of 41. 


\section{Discussion}\label{sec:discussion}

This work explored novel approaches for estimating item functioning within the \gls{glnm} framework. The study proposed two iterative procedures (a procedure using the \gls{em} algorithm and a new method based on \gls{plf}) as alternatives to the directly implemented \gls{ml} method. The methods were compared via simulation with existing algorithms and implemented in \texttt{R}.

In the simulation study, the traditional \gls{nls} approach produced the most biased parameter estimates with wide confidence intervals. The directly implemented \gls{ml} method performed satisfactorily; however, the newly proposed methods were superior in some aspects: The \gls{em} algorithm provided slightly less biased parameter estimates than the directly implemented \gls{ml} method, and it more often produced log-likelihood values closer to those computed based on true parameters. These were at the price of a higher number of iterations being needed for this approach to converge, while the maximum number of iterations was reached in several cases. As an added value, the \gls{em} algorithm provided additional information on respondents' latent response styles. The newly proposed algorithm based on \gls{plf} yielded the least biased parameter estimates of the expit function for most settings, especially when small sample sizes and additional covariates were considered. Moreover, in most scenarios, the \gls{plf}-based algorithm yielded log-likelihood values nearest to those computed based on true underlying parameters. Conversely, there was a higher rate of crashed simulations for the group-specific \gls{4pl} model \eqref{eq:4PL:dif} and small sample size. The precision of the asymptote parameters was similar for all four estimation techniques. As the sample size increased, differences between the estimation methods vanished, and all estimates were near the true values of the item parameters. 

Using two real data examples, we illustrated the possible benefits of generalized logistic regression models in item response modeling, estimating asymptotes, and their application to \gls{dif} analysis. Further, we presented how the practitioners may benefit from the added value of the \gls{em} algorithm, which can be used to estimate the probability of guessing correctly answered items (in the context of psychological assessment, endorsing an item due to pretending) or answering incorrectly due to inattention (in the context of psychological assessment, not endorsing an item due to dissimulation) for individual respondents. We also demonstrated practical challenges in estimation procedures, including specifying initial values.

\vspace{2ex}

The \gls{em} algorithm proposed in this study builds on the work of \citeA{dinse2011algorithm} while we extend their approach to the group-specific and the general covariate-specific models in multi-item measurement setting. \citeA{meng2020marginalized} proposed a similar EM algorithm for the 4PL simple model (without additional covariates) in the IRT framework. 
On the other hand, the \gls{plf}-based algorithm is novel and has not been proposed in this form for parameter estimation in the generalized logistic regression model. However, in recent decades, the idea of the \gls{plf} has been extensively discussed in the literature by many authors in various contexts, including \citeA{basu2005estimating}, \citeA{flach2014generalized}, and \citeA{scallan1984fitting}. For example, \citeA{pregibon1980goodness} proposed the \gls{ml} estimation of the link parameters using a weighted least squares algorithm. Similarly, \citeA{mccullagh1989generalized} adapted this approach and presented an algorithm in which several models with the fixed link functions were fitted. Furthermore, \citeA{kaiser1997maximum} proposed a modified scoring algorithm to perform simultaneous \gls{ml} estimation of all parameters. \citeA{scallan1984fitting} proposed an iterative two-stage algorithm, building on the work of \citeA{richards1961method}. This study examined generalized logistic regression, accounting for the possibility of guessing/pretending and inattention/dissimulation, whereas these features may depend upon the respondents' characteristics. 

The crucial part of each estimation process is specifying starting values for item parameters because these values may significantly impact the speed and precision of the estimation process. For instance, initial values far from the true item parameters may lead to situations where the estimation algorithm returns only a local extreme or does not converge. In this work, we used an approach based on an upper-lower index, resulting in low convergence rate issues with satisfactory estimation precision. However, other possible naive estimates of discrimination (and other parameters) could be considered, such as a correlation between an item score and the total test score without a given item. 

\vspace{2ex}

This study has several limitations, and several possible further directions exist. First, the simulation study was limited to two models -- the simple \gls{4pl} model \eqref{eq:4PL:simple} and the group-specific \gls{4pl} model \eqref{eq:4PL:dif}, both of which included only one or two covariates. The simulation study suggested requiring a larger sample size with an increasing number of covariates. 
Second, all the described algorithms implement the estimation item-by-item, which is typical within the GLNM framework and suitable for the cases when the ability is known or estimated apriori. The benefits include the fact that the items do not necessarily need to be independent, given the ability. One possible path for future research would be to extend the proposed algorithms to estimate all item parameters simultaneously using a joint model \cite<see, e.g.,>[Section 6.8]{martinkova2023computational}. Further, the proposed algorithms may also be implemented in the \gls{irt} framework to allow incorporating the latent ability $\theta$ and its estimation, similar to \cite{meng2020marginalized}. 
Third, this article described the \gls{nls} method as a simple approach, not accounting for the heteroscedasticity of binary data. For such data, Pearson's residuals might be more appropriate to use. This weighted form \cite<e.g.,>{ritz2015dose} takes the original squares of residuals and divides them by the variance $\pi_{pi}(1 - \pi_{pi})$. Next, the \gls{rss} of item $i$ \eqref{eq:nls:rss} would take the following form:
\begin{align*}
    \text{RSS}_i(\bgamma) = \sum_{p = 1}^n \frac{\left(Y_{pi} - \pi_{pi}\right)^2}{\pi_{pi}\left(1 - \pi_{pi}\right)}.
\end{align*}
However, the number of observations on the tails of the matching criterion is typically tiny and provides only small variability at most. These heavy weights would require a nearly exact fit for cases with few observations. Nevertheless, the computation of the \gls{nls} estimates demonstrated in this work was straightforward and efficient, providing sufficient precision. Thus, this method could be helpful in some instances, such as producing an initial idea about parameter values and using these estimates as starting values for other approaches. 
Fourth, it is important to acknowledge that the estimation methods studied here can be sensitive to the choice of optimization algorithm and the control parameters. The directly implemented \gls{ml} estimation was performed with the "L-BFGS-B" algorithm to account for constraints in asymptotes. Alternatively, asymptote parameters may depend on covariates through a transformation function, so the estimating algorithm does not need to incorporate constraints. The performance of these two approaches might differ. Moreover, the control parameters were set the same for all estimation methods, while the sensitivity of methods to their setting may vary and may be imposed in different quantities in different algorithms (e.g., deviation, likelihood, or the norm of gradient vector). For instance, the \gls{em} algorithm is known to require a large number of iterations till convergence, especially near the maximum. A potential improvement could involve a hybrid strategy, where the \gls{em} algorithm is run for a fixed number of iterations, followed by a single \gls{ml} iteration at the end. 

While the primary focus of this paper lies in enhancing parameter estimation within \gls{glnm} for multi-item measurement, it also touches upon the application of these algorithms in \gls{dif} detection. We illustrated the \gls{dif} detection by comparing the largest and the smallest models; however, a step-by-step procedure omitting the parameters with non-significant effects might be applied in practice to explore \gls{dif} in detail. Although \gls{dif} detection is not the central theme, practical examples illustrate the significance of assessing the fairness and validity of assessments across diverse groups. Nevertheless, this study does not aim to evaluate the properties of the underlying \gls{dif} detection procedure or to compare it with popular existing methods such as the anchor item-based approaches \cite{candell1988iterative, clauser1993effects, wang2003effects, kopf2015anchor} or more recent regularization based approaches \cite{magis2015detection, tutz2015penalty, belzak2020improving, wang2023using}. 

Establishing a common scale on which respondents from different groups can be scored and ranked is a crucial step in \gls{dif} analysis. In both the \gls{irt} and non-\gls{irt} frameworks (including, e.g., the Mantel-Haenszel test or SIBTEST procedure), the inclusion of \gls{dif} items in estimation or computation of ability estimate may have a severe impact on which items are detected as functioning differently. One possibility for dealing with such an issue is applying an item purification iterative algorithm \cite{lord1980applications}. Additionally, as the number of hypotheses tested may get large, p-value adjustments could be considered  \cite<see>[for discussion]{hladka2024combining}. 

In contrast to the \gls{irt} framework, \gls{glnm} offer flexibility in selecting the ability variable $\theta_p$. In this paper, the true underlying ability variable was used in the simulation study for all four estimation methods. While this approach is not feasible with real data, using a unified choice allows for a clear comparison of differences between the estimation algorithms. To account for a measurement error in total scores, one potential approach is to use plausible values; however, this may introduce greater computational complexity. Other possibilities may include the simulation-extrapolation method \cite{lockwood2014correcting, lockwood2017simulation}. In this work, we further investigated the impact of measurement error on item estimation precision through an additional simulation study (see Appendix \ref{app:sec:error}). The differences among estimation methods were similar to those observed when the true ability was used. Besides the standardized total scores, the model may utilize latent trait estimates -- possibly in an iterative algorithm, yielding an \gls{irt} model. The model may also utilize previous test scores as a matching criterion, allowing to study the differential item functioning in change \cite{martinkova2020academic}, for further applications, also see \cite{kolek2021videogames, kolek2023videogames}, or other relevant criterion variables. Additionally, the covariate-specific model \eqref{eq:4PL:gen} accommodates multi-dimensional matching criteria, similar to its \gls{irt} counterpart. Both frameworks share the same objective when accounting for the same underlying latent trait  -- to estimate item functioning with a logistic-shaped item characteristic curve. In such instances, the estimating algorithms for \gls{glnm} can provide initial estimates for the corresponding \gls{irt} model, as they are less computationally demanding, requiring lower sample size and resulting in fewer convergence issues. Moreover, they may be used for the iterative estimation of ability and item parameters. 

Finally, \gls{glnm} discussed in this paper do not account for missing data. However, when estimation and potential \gls{dif} detection are performed for each item separately, this would minimize the omission of data.

\vspace{2ex}

This study's real data examples explored item functioning in the multi-item measurement related to anxiety and learning competencies. However, the parameter estimation task in the presented models would also be relevant to many other educational, psychological, and health-related measurement areas, such as the assessment of well-being, fatigue, reading literacy, and others. Moreover, the generalized logistic regression model is not limited to multi-item measurements since the class determined by Equation \eqref{eq:4PL:gen} represents a broad family of the covariate-specific \gls{4pl} models. This model might be used and further extended in various study fields, including but not limited to quantitative pharmacology \cite{dinse2011algorithm}, applied microbiology \cite{brands2020method}, modeling patterns of urban electricity usage \cite{to2012growth}, and plant growth modeling \cite{zub2012late}. Therefore, the estimation procedures proposed in this work are highly relevant for a wide range of researchers and practitioners, both within and outside the psychometric field. 

To conclude, this study researched advances in fitting generalized logistic regression models using various estimation techniques, including two newly proposed ones. We demonstrated the superiority of the novel implementation of the \gls{em} algorithm and the newly proposed method based on \gls{plf} over the existing \gls{nls} and directly implemented \gls{ml} methods. Improving estimation algorithms is critical since it could increase precision while maintaining a user-friendly implementation. It may also provide additional information regarding individual respondents and items; thus, it is worth investing resources in the advancements of estimation methods.

\section*{Acknowledgement}

The study was funded by the Czech Science Foundation project "Theoretical Foundations of Computational Psychometrics" grant number 21-03658S, by the project "Research of Excellence on Digital Technologies and Wellbeing CZ.02.01.01/00/22\_008/0004583" which is co-financed by the European Union, and by the institutional support RVO 67985807. 
We sincerely thank the anonymous reviewers for their valuable comments and suggestions on earlier versions of the manuscript. We especially appreciate their encouragement to explore measurement errors. 
\section*{Online Supplementary Material}

Accompanying \texttt{R} scripts, simulation data, results, and figures are available at \href{https://osf.io/eu5zm/}{https://osf.io/eu5zm/}.


\bibliographystyle{apacite}
\bibliography{references}


\clearpage
\appendix
\appendixpage


\setcounter{equation}{0}
\renewcommand{\theequation}{A\arabic{equation}}


\section{Asymptotics}\label{app:sec:asymptotics}


\subsection{Nonlinear least squares}\label{app:asymptotics:nls}

Asymptotic properties of the \gls{nls} estimator, such as consistency and asymptotic distribution, can be derived under the classical set of regularity conditions \cite<e.g.,>[Theorems 5.41 and 5.42]{van2000asymptotic}. Next, one must reformulate these conditions for the covariate-specific \gls{4pl} model \eqref{eq:4PL:gen} considering $\bgammaX = (\boldsymbol{b}_i, \boldsymbol{c}_i, \boldsymbol{d}_i)$ is a vector of true parameters:
\begin{enumerate}
    \item[\namedlabel{cond:nls:R0}{[R0]}] A vector of true parameters $\bgammaX$ satisfies $$\E\left(\boldsymbol{\psi}_i(Y_{pi},\boldsymbol{X}_p; \bgammaX)\right) = \E\left(-2(Y_{pi} - \pi_{pi})\frac{\partial \pi_{pi}}{\partial \bgammaX}\right) = \boldsymbol{0}. $$
    \item[\namedlabel{cond:nls:R1}{[R1]}] The true parameter $\bgammaX$ is an interior point of the parameter space.
    \item[\namedlabel{cond:nls:R2}{[R2]}] The function $\boldsymbol{\psi}_i(y, \boldsymbol{x}; \bgamma)$ is twice continuously differentiable with respect to $\bgamma$ for every $(y, \boldsymbol{x})$.
    \item[\namedlabel{cond:nls:R3}{[R3]}] For each $\boldsymbol{\gamma}_{i}^{\ast}$ in a neighbourhood of $\bgammaX$, there exists an integrable function $\ddot{\boldsymbol{\psi}}(y, \boldsymbol{x})$ such that
    \begin{align*}
        \left|\frac{\partial^2 \psi_{ik}(y, \boldsymbol{x}; \boldsymbol{\gamma}_{i})}{\partial \gamma_{ij} \partial \gamma_{il}}\right| \leq \ddot{\boldsymbol{\psi}}(y, \boldsymbol{x}), \ \forall k, j, l.
    \end{align*}
    \item[\namedlabel{cond:nls:R4}{[R4]}] The matrix
    \begin{align*}
        \mathbb{\Gamma}_i(\bgamma) =&\ \E\left(\frac{\partial \boldsymbol{\psi}_{i}(Y_{pi}, \boldsymbol{X}_p; \bgamma)}{\partial \T{\bgamma}}\right) \\
        =&\ 2\E\left(\frac{\partial \pi_{pi}}{\partial \bgamma}\T{\left(\frac{\partial \pi_{pi}}{\partial \bgamma}\right)} - (Y_{pi} - \pi_{pi})\frac{\partial^2 \pi_{pi}}{\partial \bgamma \partial \T{\bgamma}}\right) \\
        =&\ 2\E\left(\frac{\partial \pi_{pi}}{\partial \bgamma}\T{\left(\frac{\partial \pi_{pi}}{\partial \bgamma}\right)}\right)
    \end{align*}
    is finite and regular in a neighborhood of $\bgammaX$.
    \item[\namedlabel{cond:nls:R5}{[R5]}] The variance matrix 
    \begin{align*}
        \mathbb{\Sigma}_i(\bgamma) =&\ \E\left(\boldsymbol{\psi}_{i}(Y_{pi}, \boldsymbol{X}_p; \bgamma)\T{\boldsymbol{\psi}_{i}}(Y_{pi}, \boldsymbol{X}_p; \bgamma)\right) \\
        =&\ 4\E\left(\left(Y_{pi} - \pi_{pi}\right)^2\frac{\partial \pi_{pi}}{\partial \bgamma}\T{\left(\frac{\partial \pi_{pi}}{\partial \bgamma}\right)}\right) \\
        =&\ 4\E\left(\pi_{pi} \left(1 - \pi_{pi}\right)\frac{\partial \pi_{pi}}{\partial \bgamma}\T{\left(\frac{\partial \pi_{pi}}{\partial \bgamma}\right)}\right)
    \end{align*}
    is finite for $\bgamma = \bgammaX$.
\end{enumerate}

Specifically, Theorem 5.42 \cite[p. 68]{van2000asymptotic} implies that under the conditions \ref{cond:nls:R0}--\ref{cond:nls:R5}, the probability that the estimating equations, $\frac{\partial \text{RSS}_i}{\partial \gamma_{ik}} = 0$, have at least one root tends to 1, as $n \goto \infty$, and there exists a sequence $\bgammahat$ (depending on $n$) such that $\bgammahat \gotop \bgammaX$. Moreover, the sequence $\bgammahat$ can be chosen as a local maximum for each $n$. Theorem~5.41 \cite[p. 68]{van2000asymptotic} demonstrates that every consistent estimator $\bgammahat$ has asymptotically normal distribution, that is:
\begin{align*}
    \sqrt{n} \left(\bgammahat - \bgammaX\right) \gotod \mathcal{N}\left(\boldsymbol{0}, \mathbb{\Gamma}_i^{-1}(\bgammaX) \mathbb{\Sigma}_i(\bgammaX) \mathbb{\Gamma}_i^{-1}(\bgammaX)\right),
\end{align*}

Clearly, the conditions \ref{cond:nls:R0} and \ref{cond:nls:R2} hold. To satisfy the condition \ref{cond:nls:R1}, one must bound asymptote parameters $\boldsymbol{c}_i$ and $\boldsymbol{d}_i$ to open intervals. Suppose the asymptote parameters are on the boundary of the parameter space (e.g., $c_i = 0$, $d_i = 1$, and $c_{i\text{DIF}} = d_{i\text{DIF}} = 0$ in the group-specific \gls{4pl} model \eqref{eq:4PL:dif}). In that case, the logistic regression model may be used instead. Additionally, the model \eqref{eq:4PL:dif} with~some of the parameters fixed (e.g., $d_i = 1$ and $d_{i\text{DIF}} = 0$), may be considered analogously. Notably, the asymptotic properties derived here will hold for these submodels. However, it is impossible to test whether the full model or its submodel better fits the data using, for instance, the likelihood-ratio test. Regarding the condition \ref{cond:nls:R3}, in this case, a polynomial of $x$ of the fourth degree can be taken as an integrable dominating function. 

Furthermore, because $\boldsymbol{X}_p$ describes respondent characteristics such as the standardized total score or gender, one can assume that their range is bounded, so the partial derivatives $\frac{\partial \pi_{pi}}{\partial \bgamma}$ are also bounded. Thus, matrices $\mathbb{\Gamma}_i(\bgamma)$ and $\mathbb{\Sigma}_i(\bgamma)$ are both finite, and the condition \ref{cond:nls:R5} holds. Finally, when the rows or columns of the matrix $\mathbb{\Gamma}_i(\bgamma)$ are linearly independent, the matrix has a full rank and thus is regular, satisfying condition \ref{cond:nls:R4}. For instance, the singularity of the matrix may occur when $G_p = 0, \forall p$ (or $G_p = 1, \forall p$), meaning that all respondents are from the reference (or focal) group. 

Therefore, all the assumptions \ref{cond:nls:R1}--\ref{cond:nls:R5} hold under the mild additional conditions, so $\bgammahat$ has desired asymptotic properties such as consistency and asymptotic normality of the item parameter estimates. 

\paragraph{Estimate of asymptotic variance. }
The natural estimate of the asymptotic variance of $\bgammahat$ is a \highlight{sandwich estimator} given by
\begin{align}\label{app:eq:nls:sandwich}
    \frac{1}{n} \widehat{\mathbb{\Gamma}}_{in}^{-1}(\bgammahat) \widehat{\mathbb{\Sigma}}_{in}(\bgammahat) \widehat{\mathbb{\Gamma}}_{in}^{-1}(\bgammahat),
\end{align}
where
\begin{align*}
    \widehat{\mathbb{\Gamma}}_{in}(\bgammahat) =&\ \frac{1}{n}\sum_{p = 1}^n \left(\frac{\partial \boldsymbol{\psi}_{i}(Y_{pi},\boldsymbol{X}_p; \bgammahat)}{\partial \T{\bgamma}}\right), \\
    \widehat{\mathbb{\Sigma}}_{in}(\bgamma) =&\ \frac{1}{n}\sum_{p = 1}^n \left(\boldsymbol{\psi}_{i}(Y_{pi},\boldsymbol{X}_p; \bgammahat)\T{\boldsymbol{\psi}_{i}}(Y_{pi},\boldsymbol{X}_p; \bgammahat)\right), 
\end{align*}
with components of the matrix $\widehat{\mathbb{\Gamma}}_{in}(\bgamma) = \nabla^2\text{RSS}_i(\bgamma)$ being 
\begin{align*}
    \frac{\partial^2 \text{RSS}_i(\bgamma)}{\partial \gamma_{ik} \partial \gamma_{ij}} =& \ -2 \sum_{p = 1}^n \left\{\left(Y_{pi} - \pi_{pi}\right)\frac{\partial^2 \pi_{pi}}{\partial \gamma_{ik} \partial \gamma_{ij}} -  \frac{\partial \pi_{pi}}{\partial \gamma_{ik}} \frac{\partial \pi_{pi}}{\partial \gamma_{ij}}\right\}, \\
    \frac{\partial^2 \text{RSS}_i(\bgamma)}{\partial \gamma_{ik}^2} =& \ -2 \sum_{p = 1}^n \left\{\left(Y_{pi} - \pi_{pi}\right)\frac{\partial^2 \pi_{pi}}{\partial \gamma_{ik}^2} -  \left(\frac{\partial \pi_{pi}}{\partial \gamma_{ik}}\right)^2\right\}.
\end{align*}


\subsection{Maximum likelihood}\label{app:asymptotics:ml}

Asymptotic properties of the \gls{ml} estimator can be shown under the set of the following regularity conditions \cite[Theorems 5.41 and 5.42]{van2000asymptotic}:
\begin{itemize}
    \item[\namedlabel{cond:mle:R0}{[R0$^{\ast}$]}] The support set $S = \left\{y \in \mathbb{R}: f(y|\boldsymbol{x}, \bgamma) > 0\right\}$ does not depend on the parameter $\bgamma$. 
    \item[\namedlabel{cond:mle:R1}{[R1$^{\ast}$]}] The true parameter $\bgammaX$ is an interior point of the parameter space. 
    \item[\namedlabel{cond:mle:R2}{[R2$^{\ast}$]}] The density $f(y|\boldsymbol{x}, \bgamma) = y \log\left(\pi(\boldsymbol{x}; \bgamma)\right) + (1 - y) \log\left(1 - \pi(\boldsymbol{x}; \bgamma)\right)$ is twice continuously differentiable with respect to $\bgamma$ for each $(y, \boldsymbol{x})$. 
    \item[\namedlabel{cond:mle:R3}{[R3$^{\ast}$]}] The Fisher information matrix $\mathbb{I}_{i}(\bgamma)$ is finite, regular, and positive definite in a neighborhood of $\gamma_{iX}$. 
    \item[\namedlabel{cond:mle:R4}{[R4$^{\ast}$]}] The order of differentiation and integration with respect to $\bgamma$ can be interchanged for terms $f(y|\boldsymbol{x}, \bgamma)$ and $\frac{\partial f(y|\boldsymbol{x}, \bgamma)}{\partial \bgamma}$. 
\end{itemize}

Overall, the conditions \ref{cond:mle:R0} and \ref{cond:mle:R2} hold. In this case, the regularity condition for the \gls{ml} estimator \ref{cond:mle:R1} is the same as condition \ref{cond:nls:R1} for the \gls{nls}, so one must bound parameters of asymptotes to open intervals, as discussed in Section \ref{app:asymptotics:nls}. 

For the condition \ref{cond:mle:R3}, the Fisher information matrix takes the following form:  
\begin{align*}
    \mathbb{I}_{i}(\bgamma) = \E\mathbb{I}_{i}(\bgamma|\boldsymbol{X}_p) = \E\left(\frac{1}{\pi_{pi}\left(1 - \pi_{pi}\right)}\frac{\partial \pi_{pi}}{\partial \gamma_{ik}} \frac{\partial \pi_{pi}}{\partial \gamma_{ij}}\right)_{k, j},
\end{align*}
which is a quadratic form and thus positive definite. Again, $\boldsymbol{X}_p$ describes respondent characteristics, so one can assume that their range is bounded, meaning partial derivatives $\frac{\partial \pi_{pi}}{\partial \gamma_{ik}}$, making the Fisher information matrix finite. Similarly, as in Section \ref{app:asymptotics:nls}, when the rows or columns of the Fisher information matrix $\mathbb{I}_i(\bgamma)$ are linearly independent, the matrix has a full rank and thus is regular, satisfying the condition \ref{cond:mle:R3}. The singularity of the matrix could occur in similar cases as for the matrix $\mathbb{\Gamma}_i(\bgamma)$ described in Section \ref{app:asymptotics:nls}. 

Finally, regarding the condition \ref{cond:mle:R4}, the order of differentiation and integration can be interchanged by dominated convergence theorem, as far as both $\frac{\partial f(y|\boldsymbol{x}, \bgamma)}{\partial \bgamma}$ and $\frac{\partial^2 f(y|\boldsymbol{x}, \bgamma)}{\partial \bgamma \partial \T{\bgamma}}$ are dominated by an integrable function. In this case, a polynomial of $x$ of the fourth degree can be taken as an integrable dominating function. 

\citeA{hogg2005introduction} demonstrated that when the regularity conditions \ref{cond:mle:R0}--\ref{cond:mle:R4} hold, there exists $n_0 \in \mathbf{N}$ and a sequence $\widehat{\boldsymbol{\gamma}}_{in}(n > n_0)$ of solutions to the corresponding likelihood equations such that 
\begin{align*}
    \widehat{\boldsymbol{\gamma}}_{in} \gotop \bgammaX,
\end{align*}
where $\bgammaX$ is a vector of true parameters. Because the log-likelihood function is not strictly concave, the approach described does not guarantee finding a unique solution to the corresponding likelihood equations. Thus, there might be multiple solutions, each a local maximum. However, there is one solution among them, which provides a consistent sequence of estimators. In contrast, other solutions may not even be close to $\gamma_{iX}$ and may not converge to it. Therefore, in practice, the key part of estimating procedures is finding suitable starting values, preferably easily calculated but consistent estimates of parameters. Furthermore, for this consistent sequence of solutions, it can be shown that
\begin{align*}
    \sqrt{n}\left(\widehat{\boldsymbol{\gamma}}_{in} - \bgammaX\right) \gotod \mathcal{N}(\boldsymbol{0}, \mathbb{I}^{-1}_i(\bgammaX)). 
\end{align*}

\paragraph{Estimate of asymptotic variance. }
An estimate of the asymptotic variance of the item parameters $\bgammahat$ is an inverse of the observed information matrix, an inverse of the Hessian matrix as demonstrated here:
\begin{align}\label{app:eq:mle:variance}
    \mathbb{I}_{in}^{-1}(\bgammahat|\boldsymbol{X}, \boldsymbol{G}) = \left(-\frac{1}{n} \frac{\partial^2 l_i(\bgammahat)}{\partial \bgamma \partial \T{\bgamma}}\right)^{-1}.
\end{align}


\section{Tables}\label{app:sec:tables}

\setcounter{table}{0}
\renewcommand{\thetable}{A\arabic{table}}


\begin{longtable}{lr rrr}
\caption{Item parameters estimates with standard errors in parentheses by four estimation methods using the simple \gls{4pl} model \eqref{eq:4PL:simple} for the Anxiety dataset} \label{app:tab:anxiety:pars_simple} \\
  
\toprule
Item & Method & \multicolumn{1}{c}{$b_0$} & \multicolumn{1}{c}{$b_1$} & \multicolumn{1}{c}{$d$} \\ 
\midrule 
\endfirsthead

\multicolumn{5}{c}%
{{\tablename\ \thetable{}: continued from previous page}} \\
\midrule
Item & Method & \multicolumn{1}{c}{$b_0$} & \multicolumn{1}{c}{$b_1$} & \multicolumn{1}{c}{$d$} \\ 
\midrule 
\endhead

R1 & EM & $-0.76 (-1.10, -0.42)$ & $3.01 (2.31, 3.71)$ & $0.94 (0.86, 1.01)$ \\ 
   & PLF & $-0.76 (-1.10, -0.42)$ & $3.00 (2.30, 3.70)$ & $0.94 (0.86, 1.01)$ \\ 
   \midrule 
R2 & EM & $-0.86 (-1.26, -0.46)$ & $2.89 (2.07, 3.71)$ & $0.90 (0.80, 1.00)$ \\ 
   & PLF & $-0.87 (-1.26, -0.47)$ & $2.88 (2.07, 3.70)$ & $0.90 (0.80, 1.00)$ \\ 
   \midrule 
R3 & EM & $-1.55 (-1.82, -1.28)$ & $2.69 (2.27, 3.11)$ & $1.00 (0.97, 1.03)$ \\ 
   & PLF & $-1.55 (-1.82, -1.28)$ & $2.69 (2.27, 3.11)$ & $1.00 (0.97, 1.03)$ \\ 
   \midrule 
R4 & EM & $1.17 (0.25, 2.09)$ & $3.86 (2.40, 5.33)$ & $0.98 (0.91, 1.04)$ \\ 
   & PLF & $1.15 (0.22, 2.08)$ & $3.84 (2.35, 5.33)$ & $0.98 (0.91, 1.05)$ \\ 
   \midrule 
R5 & EM & $-1.63 (-1.93, -1.34)$ & $2.87 (2.38, 3.35)$ & $0.97 (0.92, 1.02)$ \\ 
   & PLF & $-1.63 (-1.93, -1.34)$ & $2.87 (2.38, 3.35)$ & $0.97 (0.92, 1.02)$ \\ 
   \midrule 
R6 & EM & $-0.82 (-1.04, -0.59)$ & $2.42 (2.05, 2.80)$ & $1.00 (0.97, 1.03)$ \\ 
   & PLF & $-0.82 (-1.04, -0.59)$ & $2.42 (2.05, 2.80)$ & $1.00 (0.97, 1.03)$ \\ 
   \midrule 
R7 & EM & $2.40 (1.38, 3.41)$ & $4.60 (3.18, 6.02)$ & $0.97 (0.93, 1.01)$ \\ 
   & PLF & $2.39 (1.38, 3.41)$ & $4.60 (3.17, 6.02)$ & $0.97 (0.93, 1.01)$ \\ 
   \midrule 
R8 & EM & $-0.45 (-0.80, -0.09)$ & $2.05 (1.55, 2.55)$ & $0.88 (0.79, 0.98)$ \\ 
   & PLF & $-0.45 (-0.81, -0.10)$ & $2.04 (1.55, 2.54)$ & $0.88 (0.79, 0.98)$ \\ 
   \midrule 
R9 & EM & $0.41 (-0.14, 0.97)$ & $2.52 (1.75, 3.29)$ & $0.84 (0.76, 0.93)$ \\ 
   & PLF & $0.41 (-0.15, 0.96)$ & $2.51 (1.75, 3.28)$ & $0.84 (0.76, 0.93)$ \\ 
   \midrule 
R10 & EM & $-1.60 (-1.90, -1.30)$ & $3.06 (2.50, 3.63)$ & $0.98 (0.93, 1.03)$ \\ 
   & PLF & $-1.60 (-1.90, -1.30)$ & $3.06 (2.50, 3.63)$ & $0.98 (0.93, 1.03)$ \\ 
   \midrule 
R11 & EM & $0.27 (-0.20, 0.73)$ & $2.47 (1.79, 3.16)$ & $0.90 (0.82, 0.98)$ \\ 
   & PLF & $0.26 (-0.21, 0.73)$ & $2.47 (1.79, 3.15)$ & $0.90 (0.82, 0.98)$ \\ 
   \midrule 
R12 & EM & $2.00 (1.24, 2.75)$ & $4.26 (3.18, 5.35)$ & $0.93 (0.89, 0.98)$ \\ 
   & PLF & $2.00 (1.25, 2.75)$ & $4.27 (3.18, 5.35)$ & $0.93 (0.89, 0.98)$ \\ 
   \midrule 
R13 & EM & $-0.51 (-0.92, -0.09)$ & $2.03 (1.40, 2.65)$ & $0.99 (0.86, 1.11)$ \\ 
   & PLF & $-0.50 (-0.92, -0.08)$ & $2.04 (1.41, 2.67)$ & $0.98 (0.86, 1.11)$ \\ 
   \midrule 
R14 & EM & $0.34 (0.09, 0.59)$ & $2.62 (2.20, 3.05)$ & $1.00 (0.98, 1.02)$ \\ 
   & PLF & $0.34 (0.09, 0.59)$ & $2.62 (2.20, 3.05)$ & $1.00 (0.98, 1.02)$ \\ 
   \midrule 
R15 & EM & $-0.98 (-1.22, -0.74)$ & $2.24 (1.87, 2.62)$ & $1.00 (0.95, 1.05)$ \\ 
   & PLF & $-0.98 (-1.22, -0.74)$ & $2.24 (1.87, 2.62)$ & $1.00 (0.95, 1.05)$ \\ 
   \midrule 
R16 & EM & $2.49 (1.90, 3.08)$ & $4.90 (4.01, 5.80)$ & $0.98 (0.97, 1.00)$ \\ 
   & PLF & $2.49 (1.90, 3.08)$ & $4.91 (4.01, 5.80)$ & $0.98 (0.97, 1.00)$ \\ 
   \midrule 
R17 & EM & $-2.75 (-3.18, -2.32)$ & $2.54 (1.99, 3.09)$ & $0.93 (0.82, 1.04)$ \\ 
   & PLF & $-2.75 (-3.18, -2.32)$ & $2.54 (1.99, 3.09)$ & $0.93 (0.82, 1.04)$ \\ 
   \midrule 
R18 & EM & $1.51 (0.83, 2.19)$ & $3.37 (2.45, 4.29)$ & $0.90 (0.84, 0.95)$ \\ 
   & PLF & $1.51 (0.83, 2.19)$ & $3.36 (2.45, 4.28)$ & $0.90 (0.84, 0.95)$ \\ 
   \midrule 
R19 & EM & $-1.55 (-1.83, -1.27)$ & $2.80 (2.32, 3.27)$ & $0.99 (0.95, 1.03)$ \\ 
   & PLF & $-1.55 (-1.83, -1.27)$ & $2.80 (2.32, 3.28)$ & $0.99 (0.95, 1.03)$ \\ 
   \midrule 
R20 & EM & $-0.93 (-1.20, -0.67)$ & $2.83 (2.29, 3.37)$ & $0.99 (0.95, 1.04)$ \\ 
   & PLF & $-0.93 (-1.20, -0.67)$ & $2.83 (2.29, 3.38)$ & $0.99 (0.95, 1.04)$ \\ 
   \midrule 
R21 & EM & $-0.06 (-0.65, 0.53)$ & $1.95 (1.23, 2.68)$ & $0.76 (0.63, 0.89)$ \\ 
   & PLF & $-0.07 (-0.65, 0.52)$ & $1.95 (1.23, 2.66)$ & $0.76 (0.63, 0.89)$ \\ 
   \midrule 
R22 & EM & $1.16 (0.61, 1.71)$ & $4.12 (3.17, 5.08)$ & $0.98 (0.95, 1.02)$ \\ 
   & PLF & $1.16 (0.62, 1.71)$ & $4.13 (3.17, 5.08)$ & $0.98 (0.95, 1.02)$ \\ 
   \midrule 
R23 & EM & $1.06 (0.48, 1.64)$ & $3.01 (2.17, 3.85)$ & $0.96 (0.91, 1.01)$ \\ 
   & PLF & $1.06 (0.48, 1.64)$ & $3.01 (2.17, 3.85)$ & $0.96 (0.91, 1.01)$ \\ 
   \midrule 
R24 & EM & $1.52 (0.87, 2.16)$ & $4.38 (3.34, 5.42)$ & $0.93 (0.89, 0.98)$ \\ 
   & PLF & $1.51 (0.87, 2.16)$ & $4.38 (3.34, 5.41)$ & $0.93 (0.89, 0.98)$ \\ 
   \midrule 
R25 & EM & $3.30 (2.14, 4.46)$ & $4.40 (2.99, 5.81)$ & $0.93 (0.90, 0.97)$ \\ 
   & PLF & $3.29 (2.13, 4.45)$ & $4.40 (2.99, 5.81)$ & $0.93 (0.90, 0.97)$ \\ 
   \midrule 
R26 & EM & $2.55 (1.68, 3.41)$ & $4.96 (3.73, 6.20)$ & $0.95 (0.92, 0.99)$ \\ 
   & PLF & $2.55 (1.68, 3.42)$ & $4.97 (3.73, 6.20)$ & $0.95 (0.92, 0.99)$ \\ 
   \midrule 
R27 & EM & $0.93 (0.63, 1.23)$ & $3.64 (3.09, 4.19)$ & $1.00 (0.99, 1.01)$ \\ 
   & PLF & $0.93 (0.63, 1.23)$ & $3.64 (3.09, 4.19)$ & $1.00 (0.99, 1.01)$ \\ 
   \midrule 
R28 & EM & $2.60 (1.87, 3.32)$ & $5.06 (3.99, 6.14)$ & $0.98 (0.95, 1.00)$ \\ 
   & PLF & $2.59 (1.87, 3.32)$ & $5.06 (3.99, 6.13)$ & $0.98 (0.95, 1.00)$ \\ 
   \midrule 
R29 & EM & $-0.13 (-0.62, 0.36)$ & $4.08 (2.95, 5.22)$ & $0.91 (0.85, 0.98)$ \\ 
   & PLF & $-0.13 (-0.61, 0.36)$ & $4.09 (2.95, 5.23)$ & $0.91 (0.85, 0.98)$ \\ 
   \bottomrule
\end{longtable}


{\scriptsize \setlength\tabcolsep{2.5pt} 
\begin{longtable}{lr rrrrrr}
\caption{Item parameters estimates with confidence intervals in parentheses by two estimation methods using the group-specific \gls{4pl} model \eqref{eq:4PL:dif} for the Anxiety dataset} \label{app:tab:anxiety:pars_dif} \\
  
\toprule
Item & Method & \multicolumn{1}{c}{$b_0$} & \multicolumn{1}{c}{$b_1$} & \multicolumn{1}{c}{$b_2$} & \multicolumn{1}{c}{$b_3$} & \multicolumn{1}{c}{$d$} & \multicolumn{1}{c}{$d_\text{DIF}$} \\
\midrule 
\endfirsthead

\multicolumn{8}{c}%
{{\tablename\ \thetable{}: continued from previous page}} \\
\midrule
Item & Method & \multicolumn{1}{c}{$b_0$} & \multicolumn{1}{c}{$b_1$} & \multicolumn{1}{c}{$b_2$} & \multicolumn{1}{c}{$b_3$} &  \multicolumn{1}{c}{$d$} & \multicolumn{1}{c}{$d_\text{DIF}$} \\
\midrule 
\endhead
R1 & EM & $-0.73 (-1.28, -0.18)$ & $3.39 (2.25, 4.54)$ & $-0.10 (-0.83, 0.62)$ & $-0.87 (-2.38, 0.64)$ & $0.91 (0.80, 1.02)$ & $0.07 (-0.10, 0.25)$ \\ 
   & PLF & $-0.73 (-1.28, -0.18)$ & $3.39 (2.25, 4.52)$ & $-0.09 (-0.81, 0.63)$ & $-0.85 (-2.35, 0.65)$ & $0.92 (0.81, 1.02)$ & $0.07 (-0.10, 0.24)$ \\ 
   \midrule 
R2 & EM & $-0.70 (-1.53, 0.14)$ & $3.52 (1.74, 5.30)$ & $-0.29 (-1.18, 0.61)$ & $-1.16 (-3.03, 0.70)$ & $0.81 (0.65, 0.97)$ & $0.19 (0.01, 0.36)$ \\ 
   & PLF & $-0.69 (-1.53, 0.14)$ & $3.52 (1.74, 5.30)$ & $-0.29 (-1.19, 0.60)$ & $-1.17 (-3.03, 0.69)$ & $0.81 (0.65, 0.97)$ & $0.19 (0.02, 0.36)$ \\ 
   \midrule 
R3 & EM & $-1.42 (-1.88, -0.95)$ & $2.64 (1.76, 3.52)$ & $-0.23 (-0.84, 0.38)$ & $0.26 (-0.81, 1.32)$ & $0.97 (0.84, 1.10)$ & $0.03 (-0.11, 0.16)$ \\ 
   & PLF & $-1.42 (-1.88, -0.95)$ & $2.65 (1.76, 3.53)$ & $-0.23 (-0.84, 0.38)$ & $0.25 (-0.81, 1.32)$ & $0.97 (0.84, 1.10)$ & $0.03 (-0.11, 0.16)$ \\ 
   \midrule 
R4 & EM & $0.77 (0.34, 1.20)$ & $3.09 (2.37, 3.80)$ & $1.02 (-0.09, 2.13)$ & $2.12 (0.16, 4.08)$ & $1.00 (0.98, 1.02)$ & $-0.05 (-0.10, 0.01)$ \\ 
   & PLF & $0.77 (0.35, 1.20)$ & $3.08 (2.37, 3.80)$ & $1.02 (-0.09, 2.12)$ & $2.11 (0.15, 4.07)$ & $1.00 (0.98, 1.02)$ & $-0.05 (-0.10, 0.01)$ \\ 
   \midrule 
R5 & EM & $-1.67 (-2.07, -1.26)$ & $2.61 (2.02, 3.20)$ & $0.01 (-0.57, 0.60)$ & $0.47 (-0.47, 1.41)$ & $1.00 (0.95, 1.05)$ & $-0.05 (-0.14, 0.04)$ \\ 
   & PLF & $-1.67 (-2.07, -1.26)$ & $2.61 (2.02, 3.20)$ & $0.01 (-0.57, 0.60)$ & $0.47 (-0.47, 1.41)$ & $1.00 (0.95, 1.05)$ & $-0.05 (-0.14, 0.04)$ \\ 
   \midrule 
R6 & EM & $-0.46 (-0.78, -0.13)$ & $2.26 (1.76, 2.75)$ & $-0.79 (-1.28, -0.30)$ & $0.91 (-0.03, 1.84)$ & $1.00 (0.95, 1.05)$ & $-0.03 (-0.10, 0.05)$ \\ 
   & PLF & $-0.46 (-0.78, -0.13)$ & $2.26 (1.76, 2.75)$ & $-0.79 (-1.28, -0.30)$ & $0.91 (-0.02, 1.84)$ & $1.00 (0.95, 1.05)$ & $-0.03 (-0.10, 0.05)$ \\ 
   \midrule 
R7 & EM & $3.99 (2.18, 5.80)$ & $6.79 (4.30, 9.29)$ & $-2.22 (-4.10, -0.34)$ & $-3.08 (-5.70, -0.45)$ & $0.92 (0.87, 0.98)$ & $0.08 (0.02, 0.13)$ \\ 
   & PLF & $4.00 (2.19, 5.81)$ & $6.80 (4.31, 9.30)$ & $-2.22 (-4.11, -0.34)$ & $-3.08 (-5.71, -0.45)$ & $0.92 (0.87, 0.98)$ & $0.08 (0.02, 0.13)$ \\ 
   \midrule 
R8 & EM & $-0.40 (-0.96, 0.16)$ & $2.11 (1.35, 2.88)$ & $-0.09 (-0.81, 0.64)$ & $-0.13 (-1.14, 0.89)$ & $0.86 (0.72, 1.01)$ & $0.03 (-0.16, 0.22)$ \\ 
   & PLF & $-0.40 (-0.96, 0.15)$ & $2.11 (1.35, 2.87)$ & $-0.08 (-0.81, 0.64)$ & $-0.12 (-1.14, 0.89)$ & $0.87 (0.72, 1.01)$ & $0.03 (-0.16, 0.22)$ \\ 
   \midrule 
R9 & EM & $0.50 (-0.26, 1.26)$ & $2.45 (1.43, 3.47)$ & $-0.16 (-1.26, 0.93)$ & $0.28 (-1.27, 1.84)$ & $0.89 (0.77, 1.01)$ & $-0.08 (-0.25, 0.09)$ \\ 
   & PLF & $0.50 (-0.26, 1.26)$ & $2.44 (1.43, 3.46)$ & $-0.17 (-1.26, 0.93)$ & $0.28 (-1.27, 1.83)$ & $0.89 (0.77, 1.01)$ & $-0.08 (-0.25, 0.09)$ \\ 
   \midrule 
R10 & EM & $-0.98 (-1.72, -0.24)$ & $3.63 (1.76, 5.49)$ & $-1.05 (-1.92, -0.18)$ & $-0.37 (-2.38, 1.64)$ & $0.89 (0.73, 1.06)$ & $0.11 (-0.06, 0.27)$ \\ 
   & PLF & $-0.99 (-1.72, -0.26)$ & $3.61 (1.77, 5.45)$ & $-1.05 (-1.91, -0.18)$ & $-0.35 (-2.34, 1.63)$ & $0.89 (0.73, 1.06)$ & $0.11 (-0.06, 0.27)$ \\ 
   \midrule 
R11 & EM & $0.49 (-0.38, 1.35)$ & $2.72 (1.49, 3.95)$ & $-0.30 (-1.28, 0.69)$ & $-0.30 (-1.74, 1.13)$ & $0.84 (0.72, 0.97)$ & $0.09 (-0.06, 0.24)$ \\ 
   & PLF & $0.48 (-0.39, 1.34)$ & $2.71 (1.48, 3.93)$ & $-0.29 (-1.28, 0.70)$ & $-0.30 (-1.72, 1.13)$ & $0.84 (0.72, 0.97)$ & $0.09 (-0.06, 0.24)$ \\ 
   \midrule 
R12 & EM & $1.88 (1.12, 2.64)$ & $4.03 (2.91, 5.15)$ & $0.22 (-1.17, 1.61)$ & $0.55 (-1.52, 2.61)$ & $0.98 (0.94, 1.02)$ & $-0.07 (-0.14, -0.00)$ \\ 
   & PLF & $1.88 (1.12, 2.64)$ & $4.03 (2.91, 5.15)$ & $0.22 (-1.17, 1.61)$ & $0.55 (-1.52, 2.61)$ & $0.98 (0.94, 1.02)$ & $-0.07 (-0.14, -0.00)$ \\ 
   \midrule 
R13 & EM & $-0.39 (-0.75, -0.02)$ & $2.16 (1.59, 2.72)$ & $-0.22 (-0.91, 0.47)$ & $-0.19 (-1.27, 0.88)$ & $1.00 (0.94, 1.06)$ & $-0.02 (-0.23, 0.18)$ \\ 
   & PLF & $-0.39 (-0.74, -0.03)$ & $2.16 (1.60, 2.71)$ & $-0.23 (-0.92, 0.46)$ & $-0.21 (-1.28, 0.85)$ & $1.00 (0.94, 1.06)$ & $-0.02 (-0.23, 0.19)$ \\ 
   \midrule 
R14 & EM & $0.26 (-0.11, 0.63)$ & $2.54 (1.93, 3.15)$ & $0.14 (-0.36, 0.65)$ & $0.15 (-0.70, 1.00)$ & $1.00 (0.97, 1.03)$ & $0.00 (-0.04, 0.04)$ \\ 
   & PLF & $0.26 (-0.11, 0.63)$ & $2.54 (1.93, 3.15)$ & $0.14 (-0.36, 0.65)$ & $0.15 (-0.70, 1.00)$ & $1.00 (0.97, 1.03)$ & $0.00 (-0.04, 0.04)$ \\ 
   \midrule 
R15 & EM & $-1.08 (-1.45, -0.72)$ & $2.09 (1.55, 2.62)$ & $0.18 (-0.30, 0.67)$ & $0.28 (-0.46, 1.01)$ & $1.00 (0.91, 1.09)$ & $0.00 (-0.10, 0.10)$ \\ 
   & PLF & $-1.08 (-1.45, -0.72)$ & $2.09 (1.55, 2.62)$ & $0.18 (-0.30, 0.67)$ & $0.28 (-0.46, 1.01)$ & $1.00 (0.91, 1.09)$ & $0.00 (-0.10, 0.10)$ \\ 
   \midrule 
R16 & EM & $2.37 (1.56, 3.18)$ & $4.92 (3.68, 6.16)$ & $0.24 (-0.94, 1.42)$ & $-0.03 (-1.82, 1.75)$ & $0.99 (0.96, 1.01)$ & $-0.00 (-0.04, 0.04)$ \\ 
   & PLF & $2.37 (1.56, 3.18)$ & $4.92 (3.68, 6.16)$ & $0.24 (-0.94, 1.42)$ & $-0.04 (-1.82, 1.75)$ & $0.99 (0.96, 1.01)$ & $-0.00 (-0.04, 0.04)$ \\ 
   \midrule 
R17 & EM & $-3.17 (-3.98, -2.36)$ & $3.14 (2.00, 4.29)$ & $0.64 (-0.31, 1.58)$ & $-1.05 (-2.33, 0.23)$ & $0.89 (0.71, 1.06)$ & $0.11 (-0.11, 0.33)$ \\ 
   & PLF & $-3.17 (-3.98, -2.36)$ & $3.14 (2.00, 4.29)$ & $0.64 (-0.31, 1.58)$ & $-1.05 (-2.33, 0.22)$ & $0.89 (0.71, 1.06)$ & $0.11 (-0.11, 0.33)$ \\ 
   \midrule 
R18 & EM & $1.57 (0.20, 2.94)$ & $3.31 (1.57, 5.04)$ & $-0.04 (-1.65, 1.56)$ & $0.24 (-1.86, 2.33)$ & $0.88 (0.77, 1.00)$ & $0.02 (-0.11, 0.15)$ \\ 
   & PLF & $1.56 (0.22, 2.90)$ & $3.29 (1.59, 4.98)$ & $-0.03 (-1.61, 1.55)$ & $0.25 (-1.82, 2.32)$ & $0.88 (0.77, 1.00)$ & $0.02 (-0.11, 0.15)$ \\ 
   \midrule 
R19 & EM & $-1.39 (-1.77, -1.00)$ & $2.77 (2.15, 3.39)$ & $-0.33 (-0.89, 0.23)$ & $0.14 (-0.79, 1.07)$ & $1.00 (0.96, 1.04)$ & $-0.02 (-0.08, 0.05)$ \\ 
   & PLF & $-1.39 (-1.77, -1.00)$ & $2.77 (2.15, 3.39)$ & $-0.33 (-0.89, 0.23)$ & $0.14 (-0.79, 1.07)$ & $1.00 (0.96, 1.04)$ & $-0.02 (-0.08, 0.05)$ \\ 
   \midrule 
R20 & EM & $-1.09 (-1.45, -0.73)$ & $2.76 (2.13, 3.39)$ & $0.29 (-0.21, 0.79)$ & $0.12 (-0.85, 1.08)$ & $1.00 (0.96, 1.04)$ & $-0.01 (-0.08, 0.06)$ \\ 
   & PLF & $-1.09 (-1.45, -0.73)$ & $2.76 (2.13, 3.39)$ & $0.29 (-0.21, 0.79)$ & $0.11 (-0.85, 1.08)$ & $1.00 (0.96, 1.04)$ & $-0.01 (-0.08, 0.06)$ \\ 
   \midrule 
R21 & EM & $0.03 (-0.80, 0.85)$ & $1.95 (0.95, 2.95)$ & $-0.19 (-1.34, 0.96)$ & $0.13 (-1.32, 1.58)$ & $0.84 (0.66, 1.02)$ & $-0.14 (-0.39, 0.10)$ \\ 
   & PLF & $0.02 (-0.80, 0.84)$ & $1.95 (0.95, 2.94)$ & $-0.19 (-1.34, 0.95)$ & $0.12 (-1.32, 1.57)$ & $0.84 (0.66, 1.02)$ & $-0.14 (-0.39, 0.11)$ \\ 
   \midrule 
R22 & EM & $0.93 (0.45, 1.41)$ & $4.02 (3.12, 4.92)$ & $0.06 (-0.60, 0.72)$ & $-0.47 (-1.69, 0.74)$ & $1.00 (0.98, 1.02)$ & $0.00 (-0.02, 0.02)$ \\ 
   & PLF & $0.94 (0.43, 1.45)$ & $4.04 (3.08, 4.99)$ & $0.31 (-0.57, 1.19)$ & $-0.05 (-1.62, 1.51)$ & $1.00 (0.98, 1.02)$ & $-0.02 (-0.07, 0.03)$ \\ 
   \midrule 
R23 & EM & $1.34 (0.54, 2.15)$ & $3.63 (2.44, 4.81)$ & $-0.61 (-1.49, 0.27)$ & $-1.27 (-2.57, 0.04)$ & $0.94 (0.87, 1.01)$ & $0.06 (-0.01, 0.14)$ \\ 
   & PLF & $1.34 (0.54, 2.15)$ & $3.62 (2.44, 4.81)$ & $-0.61 (-1.49, 0.27)$ & $-1.27 (-2.57, 0.04)$ & $0.94 (0.87, 1.01)$ & $0.06 (-0.01, 0.14)$ \\ 
   \midrule 
R24 & EM & $1.21 (0.55, 1.87)$ & $4.11 (2.98, 5.25)$ & $0.87 (-0.49, 2.23)$ & $0.94 (-1.25, 3.14)$ & $0.98 (0.94, 1.02)$ & $-0.09 (-0.16, -0.02)$ \\ 
   & PLF & $1.21 (0.55, 1.87)$ & $4.11 (2.98, 5.25)$ & $0.87 (-0.49, 2.23)$ & $0.94 (-1.25, 3.14)$ & $0.98 (0.94, 1.02)$ & $-0.09 (-0.16, -0.02)$ \\ 
   \midrule 
R25 & EM & $3.71 (0.62, 6.80)$ & $4.83 (1.19, 8.47)$ & $-0.45 (-3.80, 2.89)$ & $-0.42 (-4.39, 3.55)$ & $0.92 (0.84, 1.00)$ & $0.02 (-0.07, 0.11)$ \\ 
   & PLF & $3.67 (0.60, 6.75)$ & $4.79 (1.17, 8.41)$ & $-0.42 (-3.75, 2.91)$ & $-0.38 (-4.33, 3.58)$ & $0.92 (0.84, 1.00)$ & $0.02 (-0.07, 0.11)$ \\ 
   \midrule 
R26 & EM & $2.87 (1.27, 4.47)$ & $5.43 (3.20, 7.66)$ & $-0.74 (-2.62, 1.14)$ & $-1.04 (-3.70, 1.62)$ & $0.97 (0.91, 1.03)$ & $-0.02 (-0.10, 0.05)$ \\ 
   & PLF & $2.86 (1.28, 4.44)$ & $5.41 (3.20, 7.63)$ & $-0.73 (-2.60, 1.13)$ & $-1.02 (-3.67, 1.62)$ & $0.97 (0.91, 1.03)$ & $-0.02 (-0.10, 0.05)$ \\ 
   \midrule 
R27 & EM & $1.08 (0.60, 1.57)$ & $3.91 (3.04, 4.79)$ & $-0.26 (-0.89, 0.36)$ & $-0.47 (-1.59, 0.66)$ & $1.00 (0.98, 1.02)$ & $0.00 (-0.02, 0.02)$ \\ 
   & PLF & $1.08 (0.60, 1.57)$ & $3.91 (3.04, 4.79)$ & $-0.26 (-0.89, 0.36)$ & $-0.47 (-1.59, 0.66)$ & $1.00 (0.98, 1.02)$ & $0.00 (-0.02, 0.02)$ \\ 
   \midrule 
R28 & EM & $2.12 (1.26, 2.98)$ & $4.54 (3.25, 5.82)$ & $0.92 (-0.44, 2.28)$ & $1.00 (-1.02, 3.02)$ & $0.98 (0.95, 1.02)$ & $-0.01 (-0.06, 0.04)$ \\ 
   & PLF & $2.12 (1.26, 2.98)$ & $4.54 (3.25, 5.83)$ & $0.92 (-0.44, 2.28)$ & $1.00 (-1.02, 3.02)$ & $0.98 (0.95, 1.02)$ & $-0.01 (-0.06, 0.04)$ \\ 
   \midrule 
R29 & EM & $0.39 (-0.21, 0.99)$ & $4.90 (3.52, 6.28)$ & $-1.07 (-1.82, -0.32)$ & $-1.69 (-3.44, 0.06)$ & $0.91 (0.84, 0.98)$ & $0.06 (-0.05, 0.18)$ \\ 
   & PLF & $0.39 (-0.21, 0.99)$ & $4.90 (3.52, 6.28)$ & $-1.07 (-1.82, -0.32)$ & $-1.68 (-3.43, 0.08)$ & $0.91 (0.84, 0.98)$ & $0.06 (-0.05, 0.17)$ \\ 
   \bottomrule
\end{longtable}
}


\begin{longtable}{lr rrr}
\caption{Item parameters estimates with standard errors in parentheses by four estimation methods using the simple \gls{4pl} model \eqref{eq:4PL:simple} for the LearningToLearn dataset} \label{app:tab:ltl:pars_simple} \\
  
\toprule
Item & Method & \multicolumn{1}{c}{$b_0$} & \multicolumn{1}{c}{$b_1$} & \multicolumn{1}{c}{$c$} \\ 
\midrule 
\endfirsthead

\multicolumn{5}{c}%
{{\tablename\ \thetable{}: continued from previous page}} \\
\midrule
Item & Method & \multicolumn{1}{c}{$b_0$} & \multicolumn{1}{c}{$b_1$} & \multicolumn{1}{c}{$d$} \\ 
\midrule 
\endhead

1A & EM & $2.01 (0.76, 3.26)$ & $1.00 (0.54, 1.46)$ & $0.01 (-1.05, 1.08)$ \\ 
   & PLF & $1.92 (0.88, 2.95)$ & $1.03 (0.61, 1.44)$ & $0.09 (-0.69, 0.88)$ \\ 
   \midrule 
1B & EM & $1.64 (0.26, 3.01)$ & $0.82 (0.48, 1.17)$ & $0.00 (-1.09, 1.09)$ \\ 
   & PLF & $1.64 (0.26, 3.01)$ & $0.82 (0.48, 1.17)$ & $0.00 (-1.09, 1.09)$ \\ 
   \midrule 
1C & EM & $1.45 (0.54, 2.36)$ & $0.74 (0.51, 0.96)$ & $0.00 (-0.69, 0.69)$ \\ 
   & PLF & $1.45 (0.54, 2.36)$ & $0.74 (0.51, 0.96)$ & $0.00 (-0.69, 0.69)$ \\ 
   \midrule 
1D & EM & $0.17 (-0.43, 0.77)$ & $0.87 (0.55, 1.18)$ & $0.00 (-0.28, 0.28)$ \\ 
   & PLF & $0.17 (-0.43, 0.77)$ & $0.87 (0.55, 1.18)$ & $0.00 (-0.28, 0.28)$ \\ 
   \midrule 
1E & EM & $-0.16 (-1.06, 0.74)$ & $0.84 (0.41, 1.26)$ & $0.17 (-0.12, 0.47)$ \\ 
   & PLF & $-0.15 (-1.05, 0.75)$ & $0.83 (0.41, 1.26)$ & $0.17 (-0.13, 0.47)$ \\ 
   \midrule 
1F & EM & $-0.02 (-0.55, 0.51)$ & $0.87 (0.61, 1.13)$ & $0.00 (-0.23, 0.23)$ \\ 
   & PLF & $-0.02 (-0.55, 0.51)$ & $0.87 (0.61, 1.13)$ & $0.00 (-0.23, 0.23)$ \\ 
   \midrule 
1G & EM & $-0.37 (-0.91, 0.16)$ & $0.87 (0.58, 1.16)$ & $0.00 (-0.18, 0.18)$ \\ 
   & PLF & $-0.37 (-0.91, 0.16)$ & $0.87 (0.58, 1.16)$ & $0.00 (-0.18, 0.18)$ \\ 
   \midrule 
1H & EM & $-1.25 (-2.18, -0.32)$ & $0.48 (0.20, 0.75)$ & $0.00 (-0.20, 0.20)$ \\ 
   & PLF & $-1.25 (-2.18, -0.32)$ & $0.48 (0.20, 0.75)$ & $0.00 (-0.20, 0.20)$ \\ 
   \midrule 
2A & EM & $3.94 (1.94, 5.93)$ & $1.11 (0.66, 1.57)$ & $0.00 (-1.82, 1.82)$ \\ 
   & PLF & $3.94 (1.94, 5.93)$ & $1.11 (0.66, 1.57)$ & $0.00 (-1.82, 1.82)$ \\ 
   \midrule 
2B & EM & $3.28 (1.46, 5.11)$ & $0.90 (0.54, 1.27)$ & $0.00 (-1.67, 1.67)$ \\ 
   & PLF & $3.28 (1.46, 5.11)$ & $0.90 (0.54, 1.27)$ & $0.00 (-1.67, 1.67)$ \\ 
   \midrule 
2C & EM & $2.91 (1.38, 4.45)$ & $1.07 (0.69, 1.45)$ & $0.00 (-1.40, 1.40)$ \\ 
   & PLF & $2.91 (1.38, 4.45)$ & $1.07 (0.69, 1.45)$ & $0.00 (-1.40, 1.40)$ \\ 
   \midrule 
2D & EM & $2.24 (1.32, 3.17)$ & $1.06 (0.77, 1.34)$ & $0.00 (-0.76, 0.76)$ \\ 
   & PLF & $2.24 (1.32, 3.17)$ & $1.06 (0.77, 1.34)$ & $0.00 (-0.76, 0.76)$ \\ 
   \midrule 
2E & EM & $1.07 (0.05, 2.08)$ & $0.73 (0.46, 0.99)$ & $0.00 (-0.70, 0.70)$ \\ 
   & PLF & $1.07 (0.05, 2.08)$ & $0.73 (0.46, 0.99)$ & $0.00 (-0.70, 0.70)$ \\ 
   \midrule 
2F & EM & $1.28 (-0.12, 2.68)$ & $0.56 (0.30, 0.82)$ & $0.00 (-1.06, 1.06)$ \\ 
   & PLF & $1.28 (-0.12, 2.68)$ & $0.56 (0.30, 0.82)$ & $0.00 (-1.06, 1.06)$ \\ 
   \midrule 
2G & EM & $-0.89 (-1.65, -0.12)$ & $0.62 (0.28, 0.96)$ & $0.00 (-0.20, 0.20)$ \\ 
   & PLF & $-0.89 (-1.66, -0.12)$ & $0.62 (0.28, 0.96)$ & $0.00 (-0.19, 0.20)$ \\ 
   \midrule 
3 & EM & $0.61 (-0.69, 1.90)$ & $0.57 (0.27, 0.87)$ & $0.00 (-0.80, 0.80)$ \\ 
   & PLF & $0.61 (-0.69, 1.90)$ & $0.57 (0.27, 0.87)$ & $0.00 (-0.80, 0.80)$ \\ 
   \midrule 
4A & EM & $-0.87 (-1.39, -0.35)$ & $1.10 (0.77, 1.43)$ & $0.00 (-0.12, 0.12)$ \\ 
   & PLF & $-0.87 (-1.39, -0.35)$ & $1.10 (0.77, 1.43)$ & $0.00 (-0.12, 0.12)$ \\ 
   \midrule 
4B & EM & $-0.53 (-1.02, -0.03)$ & $1.22 (0.89, 1.56)$ & $0.00 (-0.14, 0.14)$ \\ 
   & PLF & $-0.53 (-1.02, -0.03)$ & $1.22 (0.89, 1.56)$ & $0.00 (-0.14, 0.14)$ \\ 
   \midrule 
4C & EM & $-1.20 (-1.61, -0.78)$ & $1.39 (1.03, 1.76)$ & $0.00 (-0.05, 0.06)$ \\ 
   & PLF & $-1.20 (-1.62, -0.79)$ & $1.40 (1.03, 1.76)$ & $0.01 (-0.05, 0.06)$ \\ 
   \midrule 
4D & EM & $-1.08 (-1.34, -0.82)$ & $1.28 (1.02, 1.53)$ & $0.01 (-0.02, 0.04)$ \\ 
   & PLF & $-1.08 (-1.34, -0.82)$ & $1.28 (1.02, 1.53)$ & $0.01 (-0.02, 0.04)$ \\ 
   \midrule 
5A & EM & $1.43 (-0.13, 3.00)$ & $0.70 (0.36, 1.04)$ & $0.00 (-1.21, 1.21)$ \\ 
   & PLF & $1.43 (-0.13, 3.00)$ & $0.70 (0.36, 1.04)$ & $0.00 (-1.21, 1.21)$ \\ 
   \midrule 
5B & EM & $0.82 (-1.05, 2.69)$ & $0.41 (0.13, 0.69)$ & $0.00 (-1.27, 1.27)$ \\ 
   & PLF & $0.82 (-1.05, 2.69)$ & $0.41 (0.13, 0.69)$ & $0.00 (-1.27, 1.27)$ \\ 
   \midrule 
5C & EM & $-0.55 (-2.84, 1.74)$ & $0.40 (-0.14, 0.95)$ & $0.27 (-0.32, 0.85)$ \\ 
   & PLF & $-0.46 (-3.29, 2.37)$ & $0.38 (-0.23, 1.00)$ & $0.24 (-0.56, 1.04)$ \\ 
   \midrule 
5D & EM & $0.27 (-1.00, 1.54)$ & $0.92 (0.34, 1.50)$ & $0.35 (-0.06, 0.76)$ \\ 
   & PLF & $0.25 (-1.04, 1.53)$ & $0.93 (0.34, 1.52)$ & $0.36 (-0.05, 0.76)$ \\ 
   \midrule 
5E & EM & $1.52 (-1.93, 4.98)$ & $0.62 (0.07, 1.17)$ & $0.00 (-2.76, 2.76)$ \\ 
   & PLF & $1.52 (-2.68, 5.73)$ & $0.62 (-0.03, 1.27)$ & $0.00 (-3.36, 3.36)$ \\ 
   \midrule 
5F & EM & $-1.81 (-8.92, 5.30)$ & $0.31 (-1.39, 2.02)$ & $0.18 (-0.61, 0.97)$ \\ 
   & PLF & $-1.82 (-8.84, 5.20)$ & $0.32 (-1.38, 2.01)$ & $0.18 (-0.59, 0.95)$ \\ 
   \midrule 
5G & EM & $-5.96 (-12.03, 0.11)$ & $1.81 (-0.67, 4.29)$ & $0.11 (0.08, 0.15)$ \\ 
   & PLF & $-5.94 (-11.98, 0.10)$ & $1.80 (-0.67, 4.27)$ & $0.11 (0.08, 0.15)$ \\ 
   \midrule 
6A & EM & $-1.22 (-1.66, -0.78)$ & $1.39 (1.02, 1.75)$ & $0.03 (-0.04, 0.09)$ \\ 
   & PLF & $-1.23 (-1.67, -0.78)$ & $1.39 (1.03, 1.76)$ & $0.03 (-0.04, 0.09)$ \\ 
   \midrule 
6B & EM & $-1.47 (-2.25, -0.69)$ & $1.13 (0.67, 1.59)$ & $0.00 (-0.11, 0.11)$ \\ 
   & PLF & $-1.47 (-2.25, -0.69)$ & $1.13 (0.67, 1.59)$ & $0.00 (-0.11, 0.11)$ \\ 
   \midrule 
6C & EM & $-4.21 (-6.20, -2.22)$ & $1.60 (0.61, 2.59)$ & $0.04 (0.01, 0.07)$ \\ 
   & PLF & $-4.23 (-6.23, -2.22)$ & $1.61 (0.61, 2.60)$ & $0.04 (0.01, 0.07)$ \\ 
   \midrule 
6D & EM & $-0.60 (-1.20, -0.00)$ & $1.23 (0.79, 1.68)$ & $0.04 (-0.11, 0.19)$ \\ 
   & PLF & $-0.62 (-1.22, -0.02)$ & $1.24 (0.79, 1.69)$ & $0.04 (-0.10, 0.19)$ \\ 
   \midrule 
6E & EM & $-1.98 (-2.68, -1.28)$ & $1.23 (0.79, 1.68)$ & $0.07 (0.01, 0.13)$ \\ 
   & PLF & $-1.99 (-2.70, -1.28)$ & $1.24 (0.79, 1.68)$ & $0.07 (0.01, 0.13)$ \\ 
   \midrule 
6F & EM & $-1.63 (-2.51, -0.75)$ & $1.19 (0.65, 1.74)$ & $0.07 (-0.04, 0.17)$ \\ 
   & PLF & $-1.61 (-2.47, -0.74)$ & $1.18 (0.64, 1.71)$ & $0.06 (-0.04, 0.17)$ \\ 
   \midrule 
6G & EM & $-2.35 (-3.52, -1.18)$ & $1.03 (0.41, 1.65)$ & $0.09 (0.01, 0.16)$ \\ 
   & PLF & $-2.50 (-3.71, -1.29)$ & $1.10 (0.47, 1.74)$ & $0.10 (0.02, 0.17)$ \\ 
   \midrule 
6H & EM & $-1.69 (-3.19, -0.20)$ & $0.78 (0.09, 1.48)$ & $0.06 (-0.12, 0.25)$ \\ 
   & PLF & $-1.67 (-3.16, -0.18)$ & $0.77 (0.08, 1.46)$ & $0.06 (-0.13, 0.25)$ \\ 
   \midrule 
7A & EM & $-0.14 (-0.88, 0.60)$ & $0.92 (0.53, 1.31)$ & $0.00 (-0.29, 0.29)$ \\ 
   & PLF & $-0.14 (-0.88, 0.60)$ & $0.92 (0.53, 1.31)$ & $0.00 (-0.29, 0.29)$ \\ 
   \midrule 
7B & EM & $-0.10 (-1.27, 1.08)$ & $0.83 (0.31, 1.35)$ & $0.30 (-0.04, 0.64)$ \\ 
   & PLF & $-0.08 (-1.27, 1.11)$ & $0.82 (0.30, 1.34)$ & $0.29 (-0.06, 0.65)$ \\ 
   \midrule 
7C & EM & $-0.23 (-0.88, 0.43)$ & $0.78 (0.46, 1.10)$ & $0.00 (-0.25, 0.25)$ \\ 
   & PLF & $-0.23 (-0.88, 0.43)$ & $0.78 (0.46, 1.10)$ & $0.00 (-0.25, 0.25)$ \\ 
   \midrule 
7D & EM & $0.29 (-0.33, 0.91)$ & $0.83 (0.53, 1.14)$ & $0.00 (-0.31, 0.31)$ \\ 
   & PLF & $0.29 (-0.33, 0.91)$ & $0.83 (0.53, 1.14)$ & $0.00 (-0.31, 0.31)$ \\ 
   \midrule 
7E & EM & $-1.56 (-2.89, -0.24)$ & $0.73 (0.12, 1.35)$ & $0.03 (-0.16, 0.22)$ \\ 
   & PLF & $-1.57 (-2.89, -0.25)$ & $0.74 (0.12, 1.35)$ & $0.03 (-0.16, 0.22)$ \\ 
   \midrule 
7F & EM & $-1.09 (-2.22, 0.04)$ & $0.50 (0.14, 0.87)$ & $0.00 (-0.26, 0.26)$ \\ 
   & PLF & $-1.09 (-2.22, 0.04)$ & $0.50 (0.14, 0.87)$ & $0.00 (-0.26, 0.26)$ \\ 
   \bottomrule
\end{longtable}


\setlength\LTleft{-0.5in}
\setlength\LTright{-0.5in}
{\scriptsize \setlength\tabcolsep{3pt} 
\begin{longtable}{lr rrrrrr}
\caption{Item parameters estimates with confidence intervals in parentheses by two estimation methods using the group-specific \gls{4pl} model \eqref{eq:4PL:dif} for the LearningToLearn dataset} \label{app:tab:ltl:pars_dif} \\
  
\toprule
Item & Method & \multicolumn{1}{c}{$b_0$} & \multicolumn{1}{c}{$b_1$} & \multicolumn{1}{c}{$b_2$} & \multicolumn{1}{c}{$b_3$} & \multicolumn{1}{c}{$c$} & \multicolumn{1}{c}{$c_\text{DIF}$} \\
\midrule 
\endfirsthead

\multicolumn{8}{c}%
{{\tablename\ \thetable{}: continued from previous page}} \\
\midrule
Item & Method & \multicolumn{1}{c}{$b_0$} & \multicolumn{1}{c}{$b_1$} & \multicolumn{1}{c}{$b_2$} & \multicolumn{1}{c}{$b_3$} &  \multicolumn{1}{c}{$d$} & \multicolumn{1}{c}{$d_\text{DIF}$} \\
\midrule 
\endhead
1A & EM & $2.02 (1.15, 2.88)$ & $1.36 (0.85, 1.87)$ & $-0.94 (-3.50, 1.63)$ & $-0.50 (-1.45, 0.44)$ & $0.00 (-0.67, 0.68)$ & $0.55 (-0.45, 1.56)$ \\ 
   & PLF & $2.02 (1.15, 2.88)$ & $1.36 (0.85, 1.87)$ & $-0.92 (-3.50, 1.65)$ & $-0.51 (-1.45, 0.44)$ & $0.00 (-0.68, 0.68)$ & $0.55 (-0.47, 1.57)$ \\ 
   \midrule 
1B & EM & $1.62 (0.30, 2.94)$ & $0.85 (0.48, 1.22)$ & $-0.31 (-3.03, 2.42)$ & $0.03 (-0.81, 0.88)$ & $0.00 (-1.02, 1.02)$ & $0.23 (-1.46, 1.93)$ \\ 
   & PLF & $1.62 (0.15, 3.09)$ & $0.85 (0.46, 1.24)$ & $-0.43 (-3.11, 2.25)$ & $0.07 (-0.80, 0.93)$ & $0.00 (-1.14, 1.14)$ & $0.30 (-1.30, 1.90)$ \\ 
   \midrule 
1C & EM & $1.39 (0.07, 2.70)$ & $0.78 (0.43, 1.13)$ & $0.14 (-1.71, 1.98)$ & $-0.08 (-0.55, 0.38)$ & $0.00 (-0.98, 0.98)$ & $-0.00 (-1.39, 1.39)$ \\ 
   & PLF & $1.39 (0.07, 2.70)$ & $0.78 (0.43, 1.13)$ & $0.14 (-1.71, 1.98)$ & $-0.08 (-0.55, 0.38)$ & $0.00 (-0.98, 0.98)$ & $0.00 (-1.39, 1.39)$ \\ 
   \midrule 
1D & EM & $-0.09 (-0.74, 0.56)$ & $0.87 (0.48, 1.26)$ & $0.52 (-0.84, 1.88)$ & $0.02 (-0.67, 0.71)$ & $0.00 (-0.25, 0.25)$ & $0.01 (-0.67, 0.68)$ \\ 
   & PLF & $-0.09 (-0.74, 0.56)$ & $0.87 (0.48, 1.26)$ & $0.51 (-0.85, 1.88)$ & $0.02 (-0.67, 0.72)$ & $0.00 (-0.25, 0.25)$ & $0.01 (-0.67, 0.69)$ \\ 
   \midrule 
1E & EM & $-0.26 (-1.65, 1.12)$ & $0.83 (0.17, 1.49)$ & $0.23 (-1.61, 2.07)$ & $0.01 (-0.86, 0.88)$ & $0.13 (-0.32, 0.58)$ & $0.08 (-0.53, 0.69)$ \\ 
   & PLF & $-0.29 (-1.68, 1.09)$ & $0.85 (0.18, 1.51)$ & $0.21 (-1.64, 2.06)$ & $0.01 (-0.87, 0.90)$ & $0.14 (-0.29, 0.57)$ & $0.09 (-0.50, 0.67)$ \\ 
   \midrule 
1F & EM & $-0.14 (-0.84, 0.55)$ & $1.03 (0.64, 1.42)$ & $0.23 (-0.90, 1.37)$ & $-0.30 (-0.83, 0.22)$ & $0.00 (-0.27, 0.27)$ & $-0.00 (-0.50, 0.50)$ \\ 
   & PLF & $-0.14 (-0.84, 0.55)$ & $1.03 (0.64, 1.42)$ & $0.23 (-0.90, 1.37)$ & $-0.30 (-0.83, 0.22)$ & $0.00 (-0.27, 0.27)$ & $0.00 (-0.50, 0.50)$ \\ 
   \midrule 
1G & EM & $-0.46 (-1.20, 0.28)$ & $0.93 (0.52, 1.35)$ & $0.17 (-0.91, 1.25)$ & $-0.12 (-0.70, 0.46)$ & $0.00 (-0.23, 0.23)$ & $-0.00 (-0.37, 0.37)$ \\ 
   & PLF & $-0.46 (-1.20, 0.28)$ & $0.93 (0.52, 1.35)$ & $0.17 (-0.91, 1.25)$ & $-0.12 (-0.70, 0.46)$ & $0.00 (-0.23, 0.23)$ & $0.00 (-0.37, 0.37)$ \\ 
   \midrule 
1H & EM & $-1.39 (-3.12, 0.34)$ & $0.40 (-0.05, 0.86)$ & $0.25 (-1.77, 2.28)$ & $0.15 (-0.43, 0.72)$ & $0.00 (-0.33, 0.33)$ & $-0.00 (-0.41, 0.41)$ \\ 
   & PLF & $-1.37 (-3.10, 0.35)$ & $0.40 (-0.05, 0.85)$ & $0.24 (-1.79, 2.26)$ & $0.15 (-0.42, 0.72)$ & $0.00 (-0.34, 0.34)$ & $0.00 (-0.41, 0.41)$ \\ 
   \midrule 
2A & EM & $4.15 (-0.24, 8.54)$ & $1.08 (0.35, 1.81)$ & $-0.38 (-5.30, 4.53)$ & $0.06 (-0.88, 1.01)$ & $0.00 (-4.21, 4.21)$ & $-0.00 (-4.63, 4.63)$ \\ 
   & PLF & $4.15 (-0.24, 8.54)$ & $1.08 (0.35, 1.81)$ & $-0.38 (-5.30, 4.53)$ & $0.06 (-0.88, 1.01)$ & $0.00 (-4.21, 4.21)$ & $0.00 (-4.63, 4.63)$ \\ 
   \midrule 
2B & EM & $3.22 (-1.88, 8.32)$ & $0.67 (0.14, 1.19)$ & $0.15 (-5.24, 5.55)$ & $0.47 (-0.28, 1.21)$ & $0.00 (-4.82, 4.82)$ & $-0.00 (-5.05, 5.05)$ \\ 
   & PLF & $3.22 (-1.88, 8.32)$ & $0.67 (0.14, 1.19)$ & $0.15 (-5.24, 5.55)$ & $0.47 (-0.28, 1.21)$ & $0.00 (-4.82, 4.82)$ & $0.00 (-5.05, 5.05)$ \\ 
   \midrule 
2C & EM & $1.89 (-5.26, 9.04)$ & $0.90 (-0.70, 2.51)$ & $1.02 (-6.24, 8.28)$ & $0.45 (-1.23, 2.13)$ & $0.62 (-1.65, 2.89)$ & $-0.62 (-3.12, 1.88)$ \\ 
   & PLF & $1.95 (-5.50, 9.40)$ & $0.89 (-0.69, 2.46)$ & $0.96 (-6.60, 8.51)$ & $0.46 (-1.19, 2.11)$ & $0.60 (-1.92, 3.12)$ & $-0.60 (-3.33, 2.13)$ \\ 
   \midrule 
2D & EM & $2.28 (0.76, 3.79)$ & $1.21 (0.66, 1.76)$ & $-0.06 (-2.08, 1.96)$ & $-0.31 (-0.97, 0.35)$ & $0.00 (-1.29, 1.29)$ & $-0.00 (-1.69, 1.69)$ \\ 
   & PLF & $2.28 (0.76, 3.79)$ & $1.21 (0.66, 1.76)$ & $-0.06 (-2.08, 1.96)$ & $-0.31 (-0.97, 0.35)$ & $0.00 (-1.29, 1.29)$ & $0.00 (-1.69, 1.69)$ \\ 
   \midrule 
2E & EM & $1.03 (-0.50, 2.56)$ & $0.78 (0.33, 1.23)$ & $0.07 (-2.01, 2.15)$ & $-0.11 (-0.67, 0.46)$ & $0.00 (-1.05, 1.05)$ & $-0.00 (-1.44, 1.44)$ \\ 
   & PLF & $1.03 (-0.50, 2.56)$ & $0.78 (0.33, 1.23)$ & $0.07 (-2.01, 2.15)$ & $-0.11 (-0.67, 0.46)$ & $0.00 (-1.05, 1.05)$ & $0.00 (-1.44, 1.44)$ \\ 
   \midrule 
2F & EM & $0.54 (-1.61, 2.68)$ & $0.70 (-0.02, 1.42)$ & $0.66 (-1.84, 3.16)$ & $-0.10 (-0.87, 0.67)$ & $0.44 (-0.27, 1.14)$ & $-0.44 (-1.61, 0.73)$ \\ 
   & PLF & $0.50 (-1.59, 2.59)$ & $0.71 (-0.00, 1.43)$ & $0.69 (-1.75, 3.14)$ & $-0.11 (-0.88, 0.66)$ & $0.45 (-0.21, 1.11)$ & $-0.45 (-1.59, 0.69)$ \\ 
   \midrule 
2G & EM & $-1.07 (-2.02, -0.11)$ & $0.72 (0.25, 1.20)$ & $0.19 (-1.51, 1.89)$ & $-0.14 (-0.84, 0.56)$ & $0.04 (-0.16, 0.24)$ & $-0.04 (-0.46, 0.39)$ \\ 
   & PLF & $-1.09 (-2.06, -0.11)$ & $0.73 (0.25, 1.21)$ & $0.21 (-1.50, 1.92)$ & $-0.15 (-0.85, 0.56)$ & $0.04 (-0.16, 0.24)$ & $-0.04 (-0.47, 0.38)$ \\ 
   \midrule 
3 & EM & $0.78 (-0.32, 1.88)$ & $0.67 (0.37, 0.97)$ & $-0.85 (-3.57, 1.87)$ & $-0.04 (-0.91, 0.83)$ & $0.00 (-0.70, 0.70)$ & $0.23 (-0.86, 1.33)$ \\ 
   & PLF & $0.78 (-0.49, 2.05)$ & $0.67 (0.35, 0.98)$ & $-0.78 (-3.75, 2.18)$ & $-0.06 (-0.96, 0.83)$ & $0.00 (-0.81, 0.81)$ & $0.21 (-1.06, 1.48)$ \\ 
   \midrule 
4A & EM & $-0.95 (-1.75, -0.15)$ & $1.04 (0.57, 1.52)$ & $0.16 (-0.89, 1.21)$ & $0.12 (-0.54, 0.78)$ & $0.00 (-0.18, 0.18)$ & $-0.00 (-0.24, 0.24)$ \\ 
   & PLF & $-0.95 (-1.75, -0.15)$ & $1.04 (0.57, 1.52)$ & $0.16 (-0.89, 1.21)$ & $0.12 (-0.54, 0.78)$ & $0.00 (-0.18, 0.18)$ & $0.00 (-0.24, 0.24)$ \\ 
   \midrule 
4B & EM & $-0.62 (-1.35, 0.12)$ & $1.12 (0.67, 1.57)$ & $0.18 (-0.81, 1.17)$ & $0.21 (-0.47, 0.89)$ & $0.00 (-0.20, 0.20)$ & $-0.00 (-0.28, 0.28)$ \\ 
   & PLF & $-0.62 (-1.35, 0.12)$ & $1.12 (0.67, 1.57)$ & $0.18 (-0.81, 1.17)$ & $0.21 (-0.47, 0.89)$ & $0.00 (-0.20, 0.20)$ & $0.00 (-0.28, 0.28)$ \\ 
   \midrule 
4C & EM & $-1.53 (-2.11, -0.96)$ & $1.69 (1.14, 2.23)$ & $0.51 (-0.41, 1.44)$ & $-0.44 (-1.19, 0.31)$ & $0.02 (-0.03, 0.08)$ & $-0.02 (-0.17, 0.13)$ \\ 
   & PLF & $-1.54 (-2.12, -0.96)$ & $1.69 (1.15, 2.23)$ & $0.52 (-0.40, 1.44)$ & $-0.44 (-1.19, 0.31)$ & $0.02 (-0.03, 0.08)$ & $-0.02 (-0.17, 0.13)$ \\ 
   \midrule 
4D & EM & $-1.25 (-1.93, -0.57)$ & $1.40 (0.91, 1.89)$ & $0.30 (-0.49, 1.09)$ & $-0.22 (-0.83, 0.39)$ & $0.00 (-0.11, 0.11)$ & $0.02 (-0.11, 0.15)$ \\ 
   & PLF & $-1.25 (-1.93, -0.57)$ & $1.40 (0.91, 1.89)$ & $0.30 (-0.50, 1.09)$ & $-0.22 (-0.83, 0.39)$ & $0.00 (-0.11, 0.11)$ & $0.02 (-0.11, 0.15)$ \\ 
   \midrule 
5A & EM & $1.41 (-0.98, 3.80)$ & $0.76 (0.15, 1.38)$ & $-0.10 (-3.15, 2.94)$ & $-0.08 (-0.82, 0.65)$ & $0.12 (-1.50, 1.74)$ & $-0.12 (-2.26, 2.03)$ \\ 
   & PLF & $1.41 (-1.01, 3.83)$ & $0.76 (0.14, 1.39)$ & $-0.10 (-3.17, 2.97)$ & $-0.08 (-0.82, 0.66)$ & $0.12 (-1.51, 1.75)$ & $-0.12 (-2.28, 2.04)$ \\ 
   \midrule 
5B & EM & $0.83 (-1.46, 3.13)$ & $0.41 (0.08, 0.75)$ & $-0.24 (-3.62, 3.14)$ & $0.02 (-0.56, 0.60)$ & $0.00 (-1.56, 1.56)$ & $0.13 (-1.93, 2.19)$ \\ 
   & PLF & $0.83 (-1.47, 3.14)$ & $0.41 (0.07, 0.75)$ & $-0.25 (-3.80, 3.29)$ & $0.02 (-0.59, 0.63)$ & $0.00 (-1.57, 1.57)$ & $0.14 (-1.99, 2.26)$ \\ 
   \midrule 
5C & EM & $0.14 (-1.79, 2.07)$ & $0.44 (0.02, 0.87)$ & $-2.42 (-7.20, 2.35)$ & $0.24 (-1.56, 2.04)$ & $0.00 (-0.98, 0.99)$ & $0.48 (-0.52, 1.48)$ \\ 
   & PLF & $0.15 (-1.80, 2.09)$ & $0.44 (0.01, 0.87)$ & $-2.43 (-7.21, 2.35)$ & $0.24 (-1.56, 2.04)$ & $0.00 (-1.00, 1.00)$ & $0.48 (-0.53, 1.50)$ \\ 
   \midrule 
5D & EM & $-0.13 (-1.68, 1.42)$ & $1.25 (0.22, 2.28)$ & $0.99 (-2.21, 4.19)$ & $-0.59 (-1.85, 0.66)$ & $0.45 (0.15, 0.76)$ & $-0.38 (-2.14, 1.38)$ \\ 
   & PLF & $-0.11 (-1.66, 1.44)$ & $1.24 (0.21, 2.27)$ & $0.97 (-2.20, 4.14)$ & $-0.58 (-1.83, 0.66)$ & $0.45 (0.13, 0.77)$ & $-0.38 (-2.12, 1.36)$ \\ 
   \midrule 
5E & EM & $-0.21 (-1.74, 1.33)$ & $2.21 (0.31, 4.10)$ & $1.52 (-2.69, 5.73)$ & $-1.77 (-3.72, 0.18)$ & $0.67 (0.52, 0.81)$ & $-0.63 (-3.55, 2.29)$ \\ 
   & PLF & $-0.19 (-1.73, 1.35)$ & $2.19 (0.30, 4.08)$ & $1.50 (-3.74, 6.75)$ & $-1.75 (-3.72, 0.23)$ & $0.67 (0.52, 0.82)$ & $-0.63 (-4.36, 3.10)$ \\ 
   \midrule 
5F & EM & $-1.33 (-25.85, 23.19)$ & $0.08 (-1.38, 1.54)$ & $-0.86 (-25.71, 23.99)$ & $0.55 (-1.63, 2.74)$ & $0.13 (-4.34, 4.60)$ & $0.07 (-4.41, 4.55)$ \\ 
   & PLF & $-1.35 (-31.95, 29.25)$ & $0.08 (-1.76, 1.92)$ & $-0.79 (-31.68, 30.09)$ & $0.54 (-1.94, 3.01)$ & $0.13 (-5.34, 5.60)$ & $0.06 (-5.42, 5.54)$ \\ 
   \midrule 
5G & EM & $-2.38 (-6.82, 2.07)$ & $0.53 (-1.11, 2.17)$ & $-4.96 (-12.91, 2.99)$ & $1.69 (-1.47, 4.84)$ & $0.05 (-0.28, 0.38)$ & $0.05 (-0.27, 0.38)$ \\ 
   & PLF & $-2.37 (-6.79, 2.05)$ & $0.53 (-1.10, 2.16)$ & $-4.98 (-12.93, 2.98)$ & $1.69 (-1.47, 4.85)$ & $0.05 (-0.28, 0.38)$ & $0.05 (-0.27, 0.38)$ \\ 
   \midrule 
6A & EM & $-1.22 (-1.77, -0.67)$ & $1.43 (0.96, 1.91)$ & $0.11 (-1.05, 1.27)$ & $-0.17 (-1.05, 0.70)$ & $0.03 (-0.04, 0.11)$ & $-0.03 (-0.23, 0.16)$ \\ 
   & PLF & $-1.22 (-1.78, -0.67)$ & $1.44 (0.96, 1.91)$ & $0.12 (-1.05, 1.29)$ & $-0.18 (-1.06, 0.70)$ & $0.04 (-0.04, 0.11)$ & $-0.04 (-0.23, 0.16)$ \\ 
   \midrule 
6B & EM & $-1.41 (-2.27, -0.55)$ & $1.19 (0.65, 1.73)$ & $-0.13 (-2.02, 1.77)$ & $-0.12 (-1.21, 0.98)$ & $0.00 (-0.13, 0.13)$ & $-0.00 (-0.27, 0.27)$ \\ 
   & PLF & $-1.41 (-2.27, -0.55)$ & $1.19 (0.65, 1.73)$ & $-0.13 (-2.02, 1.77)$ & $-0.12 (-1.21, 0.98)$ & $0.00 (-0.13, 0.13)$ & $0.00 (-0.27, 0.27)$ \\ 
   \midrule 
6C & EM & $-5.21 (-9.61, -0.81)$ & $2.08 (0.00, 4.15)$ & $1.52 (-3.36, 6.40)$ & $-0.74 (-3.09, 1.61)$ & $0.06 (0.02, 0.10)$ & $-0.04 (-0.10, 0.02)$ \\ 
   & PLF & $-5.21 (-9.61, -0.81)$ & $2.08 (0.00, 4.15)$ & $1.55 (-3.32, 6.42)$ & $-0.75 (-3.09, 1.59)$ & $0.06 (0.02, 0.10)$ & $-0.04 (-0.10, 0.02)$ \\ 
   \midrule 
6D & EM & $-0.59 (-1.79, 0.60)$ & $1.14 (0.37, 1.92)$ & $-0.07 (-1.42, 1.28)$ & $0.21 (-0.75, 1.16)$ & $0.07 (-0.23, 0.38)$ & $-0.06 (-0.39, 0.27)$ \\ 
   & PLF & $-0.54 (-1.72, 0.65)$ & $1.11 (0.35, 1.87)$ & $-0.12 (-1.47, 1.22)$ & $0.24 (-0.70, 1.18)$ & $0.06 (-0.26, 0.38)$ & $-0.05 (-0.39, 0.30)$ \\ 
   \midrule 
6E & EM & $-1.74 (-2.84, -0.65)$ & $1.03 (0.39, 1.66)$ & $-0.43 (-1.84, 0.98)$ & $0.39 (-0.48, 1.27)$ & $0.09 (-0.03, 0.21)$ & $-0.03 (-0.17, 0.10)$ \\ 
   & PLF & $-1.77 (-2.87, -0.67)$ & $1.04 (0.40, 1.68)$ & $-0.41 (-1.82, 1.01)$ & $0.38 (-0.50, 1.26)$ & $0.09 (-0.03, 0.21)$ & $-0.04 (-0.17, 0.10)$ \\ 
   \midrule 
6F & EM & $-2.56 (-4.12, -1.01)$ & $1.77 (0.72, 2.81)$ & $1.18 (-0.59, 2.95)$ & $-0.67 (-1.83, 0.49)$ & $0.16 (0.08, 0.24)$ & $-0.16 (-0.32, -0.01)$ \\ 
   & PLF & $-2.55 (-4.11, -1.00)$ & $1.76 (0.72, 2.80)$ & $1.17 (-0.60, 2.93)$ & $-0.66 (-1.82, 0.49)$ & $0.16 (0.08, 0.24)$ & $-0.16 (-0.32, -0.01)$ \\ 
   \midrule 
6G & EM & $-2.70 (-4.29, -1.11)$ & $1.35 (0.46, 2.24)$ & $0.76 (-1.82, 3.33)$ & $-0.63 (-1.91, 0.66)$ & $0.13 (0.05, 0.20)$ & $-0.10 (-0.32, 0.13)$ \\ 
   & PLF & $-2.69 (-4.26, -1.11)$ & $1.34 (0.46, 2.23)$ & $0.82 (-1.75, 3.38)$ & $-0.65 (-1.92, 0.62)$ & $0.12 (0.05, 0.20)$ & $-0.10 (-0.34, 0.14)$ \\ 
   \midrule 
6H & EM & $-16.03 (-35.65, 3.59)$ & $7.34 (-1.57, 16.26)$ & $14.46 (-5.18, 34.11)$ & $-6.49 (-15.42, 2.44)$ & $0.23 (0.18, 0.27)$ & $-0.21 (-0.36, -0.06)$ \\ 
   & PLF & $-16.03 (-35.65, 3.59)$ & $7.34 (-1.57, 16.26)$ & $14.46 (-5.19, 34.11)$ & $-6.49 (-15.42, 2.44)$ & $0.23 (0.18, 0.27)$ & $-0.21 (-0.35, -0.06)$ \\ 
   \midrule 
7A & EM & $-0.01 (-1.18, 1.16)$ & $0.94 (0.31, 1.58)$ & $-0.29 (-1.81, 1.22)$ & $-0.02 (-0.83, 0.78)$ & $0.01 (-0.47, 0.49)$ & $-0.01 (-0.60, 0.58)$ \\ 
   & PLF & $-0.01 (-1.18, 1.17)$ & $0.94 (0.31, 1.58)$ & $-0.29 (-1.81, 1.22)$ & $-0.02 (-0.83, 0.78)$ & $0.01 (-0.47, 0.49)$ & $-0.01 (-0.60, 0.58)$ \\ 
   \midrule 
7B & EM & $0.65 (-0.92, 2.23)$ & $0.52 (0.19, 0.84)$ & $-1.68 (-3.68, 0.31)$ & $1.13 (0.12, 2.15)$ & $0.00 (-0.99, 0.99)$ & $0.43 (-0.57, 1.43)$ \\ 
   & PLF & $0.66 (-0.92, 2.24)$ & $0.52 (0.19, 0.84)$ & $-1.68 (-3.68, 0.31)$ & $1.13 (0.12, 2.15)$ & $0.00 (-0.99, 0.99)$ & $0.43 (-0.57, 1.43)$ \\ 
   \midrule 
7C & EM & $-0.32 (-1.40, 0.76)$ & $0.62 (0.21, 1.03)$ & $0.07 (-1.37, 1.51)$ & $0.42 (-0.31, 1.14)$ & $0.00 (-0.41, 0.41)$ & $0.04 (-0.48, 0.56)$ \\ 
   & PLF & $-0.32 (-1.39, 0.76)$ & $0.62 (0.21, 1.03)$ & $0.03 (-1.40, 1.46)$ & $0.44 (-0.28, 1.17)$ & $0.00 (-0.41, 0.41)$ & $0.06 (-0.45, 0.57)$ \\ 
   \midrule 
7D & EM & $-0.11 (-1.18, 0.96)$ & $0.90 (0.34, 1.47)$ & $0.54 (-0.73, 1.81)$ & $-0.00 (-0.68, 0.67)$ & $0.12 (-0.26, 0.49)$ & $-0.12 (-0.63, 0.40)$ \\ 
   & PLF & $-0.10 (-1.16, 0.96)$ & $0.90 (0.34, 1.45)$ & $0.53 (-0.74, 1.79)$ & $0.00 (-0.66, 0.67)$ & $0.11 (-0.27, 0.49)$ & $-0.11 (-0.63, 0.41)$ \\ 
   \midrule 
7E & EM & $-1.32 (-2.82, 0.18)$ & $0.54 (0.02, 1.06)$ & $-1.18 (-3.22, 0.86)$ & $0.82 (-0.16, 1.81)$ & $0.00 (-0.29, 0.29)$ & $0.09 (-0.21, 0.40)$ \\ 
   & PLF & $-1.31 (-2.81, 0.19)$ & $0.54 (0.02, 1.06)$ & $-1.17 (-3.21, 0.87)$ & $0.82 (-0.16, 1.80)$ & $0.00 (-0.29, 0.29)$ & $0.09 (-0.21, 0.40)$ \\ 
   \midrule 
7F & EM & $-0.99 (-2.43, 0.44)$ & $0.45 (0.07, 0.82)$ & $-0.50 (-2.66, 1.67)$ & $0.24 (-0.58, 1.06)$ & $0.00 (-0.37, 0.37)$ & $0.05 (-0.39, 0.50)$ \\ 
   & PLF & $-0.99 (-2.43, 0.44)$ & $0.45 (0.07, 0.82)$ & $-0.49 (-2.66, 1.69)$ & $0.24 (-0.59, 1.06)$ & $0.00 (-0.37, 0.37)$ & $0.05 (-0.40, 0.50)$ \\ 
   \bottomrule
\end{longtable}
}


\section{Measurement error}\label{app:sec:error}

To assess the influence of measurement error on the estimation techniques, we performed an additional simulation study. The simple model \eqref{eq:4PL:simple} was considered only. 

\subsection{Simulation design}

\paragraph{Data generation.} To generate data, one set of item parameters was considered with the same values as in the main simulation study. Two matching criteria were considered: First, the true ability $\theta_p$ was generated from the standard normal distribution $\mathcal{N}(0, 1)$; second, we generated $\theta_p^* = \theta_p + e_p$, where $e_p \sim \mathcal{N}(0, \sqrt{0.1})$ to ensure that correlation between these two matching criteria is equal to 0.9. Two sets of binary responses were generated from the Bernoulli distribution with the calculated probabilities based on the simple \gls{4pl} model \eqref{eq:4PL:simple}, true parameters, and the
matching criterion variable. The sample size was set to $n =$ 500; 1,000; 2,500; and 5,000. Each scenario was replicated 1,000 times.

\paragraph{Simulation evaluation. } To assess the impact of the measurement error on estimation precision in item parameters for all four estimation algorithms, we computed the bias of item parameters for both sets of binary responses. Analogously to the main simulation study, we selected only those simulation runs for which all four estimation methods converged successfully and where the absolute value of the item parameter estimates did not exceed 100. 

\subsection{Simulation results}

For the parameters $b_0$, $c$, and $d$, the bias was decreasing with the increasing sample size for both ability variables. With a sample size of $n = 500$, the bias was higher for these item parameter estimates based on the ability variable $\theta_p^*$, which incorporates a measurement error, but was comparable to estimates based on true ability $\theta_p$ at larger sample sizes (Table~\ref{app:tab:results}). For the slope parameter $b_1$, the situation was reversed. At a sample size of $n = 500$, the bias was smaller when using $\theta_p^*$. However, it was not decreasing with the increasing sample size, as it was when the true ability $\theta_p$ was considered.

The most precise estimates of the intercept parameter $b_0$ were obtained by the \gls{nls} method, followed by the \gls{plf}-based algorithm. The least biased estimates of the slope parameter $b_1$ were produced by the \gls{plf}-based algorithm. The precision for the asymptote parameters $c$ and $d$ was comparable across all four estimation approaches. As the sample size increased, the differences between methods diminished.

\begin{table}[ht]
\centering
\caption{Bias of item parameters $b_0$, $b_1$, $c$, and $d$ with respect to estimation method (\gls{nls}, \gls{ml}, \gls{em} algorithm, and \gls{plf}-based algorithm), sample size $n$, and the choice of the ability variable, either $\theta_p$ pr $\theta_p^*$. }\label{app:tab:results} 
\begin{tabular}{r rrrrrrrr}
  \toprule
\multirow{2}{*}{Method / $n$} & \multicolumn{2}{c}{$b_0$} & \multicolumn{2}{c}{$b_1$} & 
\multicolumn{2}{c}{$c$} & \multicolumn{2}{c}{$d$}\\ \cmidrule(lr){2-3} \cmidrule(lr){4-5} \cmidrule(lr){6-7}
\cmidrule(lr){8-9}
& \multicolumn{1}{c}{$\theta_p$} & \multicolumn{1}{c}{$\theta_p^*$}
& \multicolumn{1}{c}{$\theta_p$} & \multicolumn{1}{c}{$\theta_p^*$}
& \multicolumn{1}{c}{$\theta_p$} & \multicolumn{1}{c}{$\theta_p^*$}
& \multicolumn{1}{c}{$\theta_p$} & \multicolumn{1}{c}{$\theta_p^*$}\\
  \midrule
  \multicolumn{9}{l}{NLS} \\
 \hspace{1em}500 & $0.009$ & $-0.017$ & $0.677$ & $0.634$ & $0.007$ & $0.011$ & $-0.012$ & $-0.016$ \\ 
   \hspace{1em}1,000 & $-0.018$ & $-0.013$ & $0.271$ & $0.071$ & $0.006$ & $0.007$ & $-0.005$ & $-0.005$ \\ 
   \hspace{1em}2,500 & $0.002$ & $0.006$ & $0.140$ & $-0.079$ & $0.006$ & $0.003$ & $-0.006$ & $-0.004$ \\ 
   \hspace{1em}5,000 & $0.000$ & $-0.000$ & $0.052$ & $-0.148$ & $0.001$ & $0.000$ & $-0.001$ & $-0.000$ \\ 
   \multicolumn{9}{l}{ML} \\
   \hspace{1em}500 & $-0.041$ & $-0.065$ & $0.668$ & $0.609$ & $0.009$ & $0.013$ & $-0.008$ & $-0.012$ \\ 
 \hspace{1em}1,000 & $-0.031$ & $-0.029$ & $0.264$ & $0.062$ & $0.008$ & $0.008$ & $-0.004$ & $-0.005$ \\ 
  \hspace{1em}2,500 & $-0.002$ & $0.001$ & $0.136$ & $-0.084$ & $0.006$ & $0.004$ & $-0.006$ & $-0.004$ \\ 
  \hspace{1em}5,000 & $-0.000$ & $-0.002$ & $0.051$ & $-0.149$ & $0.002$ & $0.001$ & $-0.001$ & $-0.000$ \\
  \multicolumn{9}{l}{EM} \\
  \hspace{1em}500 & $-0.039$ & $-0.068$ & $0.661$ & $0.586$ & $0.011$ & $0.016$ & $-0.010$ & $-0.014$ \\ 
  \hspace{1em}1,000 & $-0.034$ & $-0.034$ & $0.269$ & $0.082$ & $0.010$ & $0.013$ & $-0.005$ & $-0.008$ \\ 
  \hspace{1em}2,500 & $-0.005$ & $-0.004$ & $0.146$ & $-0.059$ & $0.008$ & $0.008$ & $-0.007$ & $-0.008$ \\ 
  \hspace{1em}5,000 & $-0.004$ & $-0.005$ & $0.064$ & $-0.127$ & $0.004$ & $0.004$ & $-0.003$ & $-0.003$ \\ 
  \multicolumn{9}{l}{PLF} \\
  \hspace{1em}500 & $-0.026$ & $-0.039$ & $0.486$ & $0.256$ & $0.006$ & $0.010$ & $-0.006$ & $-0.010$ \\ 
  \hspace{1em}1,000 & $-0.028$ & $-0.029$ & $0.224$ & $0.040$ & $0.008$ & $0.009$ & $-0.004$ & $-0.005$ \\ 
  \hspace{1em}2,500 & $-0.003$ & $-0.001$ & $0.145$ & $-0.069$ & $0.008$ & $0.007$ & $-0.007$ & $-0.007$ \\ 
  \hspace{1em}5,000 & $-0.002$ & $-0.004$ & $0.071$ & $-0.126$ & $0.004$ & $0.005$ & $-0.004$ & $-0.004$ \\ 
   \bottomrule
\end{tabular}
\end{table}


\end{document}